\documentclass[aps,pra,twocolumn,preprintnumbers,superscriptaddress,amsmath,amssymb,footinbib,longbibliography]{revtex4-2}
\usepackage{graphicx,epsfig}
\usepackage{bm}
\usepackage{dcolumn}
\usepackage[english]{babel}
\usepackage[utf8]{inputenc}
\usepackage{comment}
\usepackage{amsmath}
\usepackage{scalerel}
\usepackage{lipsum}
\usepackage{amsfonts}
\usepackage{amssymb}
\usepackage{color}
\usepackage[usenames,dvipsnames]{xcolor}
\usepackage[colorlinks,bookmarks=false,citecolor=NavyBlue,linkcolor=Red,urlcolor=blue,
]{hyperref}
\usepackage[export]{adjustbox}
\usepackage{amsthm}
\usepackage{simplewick}
\usepackage{bbold}
\usepackage{ wasysym }
\usepackage[title]{appendix}
\usepackage{mathrsfs}  
\usepackage{braket}
\usepackage{graphics}
\usepackage{todonotes}
\usepackage{subfigure}

\usepackage{multirow}
\usepackage{algorithm}
\usepackage{algpseudocode}

\usepackage{enumitem}

\newcommand{\be}{\begin{equation}}
\newcommand{\ee}{\end{equation}}
\newcommand{\bea}{\begin{eqnarray}}
\newcommand{\eea}{\end{eqnarray}}
\newcommand{\llangle}{\langle\!\langle}
\newcommand{\rrangle}{\rangle\!\rangle}
\newcommand{\tr}{\text{tr}}
\newcommand{\norm}[1]{\lVert{#1}\rVert}

\begin{document}

%
\title{Accuracy of time-dependent GGE under weak dissipation}

\author{Luca Lumia}
\affiliation{International School for Advanced Studies (SISSA), 34136 Trieste, Italy}

\author{Gianni Aupetit-Diallo}
\affiliation{International School for Advanced Studies (SISSA), 34136 Trieste, Italy}

\author{J\'er\^ome Dubail}
\affiliation{CESQ and ISIS (UMR 7006), University of Strasbourg and CNRS, 67000 Strasbourg, France}

\author{Mario Collura}
\affiliation{International School for Advanced Studies (SISSA), 34136 Trieste, Italy}
\affiliation{INFN Sezione di Trieste, 34136 Trieste, Italy}

\begin{abstract}

Unitary integrable models typically relax to a stationary Generalized Gibbs Ensemble (GGE), but in experimental realizations dissipation often breaks integrability. In this work, we use the recently introduced time-dependent GGE (t-GGE) approach to describe the open dynamics of a gas of bosons subject to atom losses and gains. We employ tensor network methods to provide numerical evidence of the exactness of the t-GGE in the limit of adiabatic dissipation,  and of its accuracy in the regime of weak but finite dissipation. That accuracy is tested for two-point functions via the rapidity distribution, and for more complicated correlations through a non-Gaussianity measure.
We combine this description with Generalized Hydrodynamics and we show that it correctly captures transport at large scales.
Our results demonstrate that the t-GGE approach is robust in both homogeneous and inhomogeneous settings.
\end{abstract}


\maketitle

\section{Introduction}

The difficulty to accurately simulate the complex dynamics of many-body quantum systems coupled to a dissipative environment~\cite{rivas2012open,breuer2002theory} is currently a bottleneck across various fields of research. These include quantum condensed matter and cold atom platforms, where a fundamental motivation is that quantum systems in nature are never perfectly isolated but are constantly interacting with their environment via heat transfer, decoherence, atom losses, etc. \cite{schlosshauer2004decoherence,vinjanampathy2016quantum,weiner1999experiments,bouchoule2022generalized}; another important motivation is the envisioned possibility to use controlled dissipation channels to engineer interesting quantum many-body stationary states, sometimes with features not found in isolated systems~\cite{diehl2008quantum,verstraete2009quantum}. Dissipative quantum many-body dynamics is also a central research topic in quantum information~\cite{harrington2022engineered}, where dissipation is essential to such basic tasks as qubit resetting and measurement, and where it finds applications in quantum error correction~\cite{cohen2014dissipation,lebreuilly2021autonomous} or quantum sensing~\cite{reiter2017dissipative}. It is also a central topic in quantum chemistry~\cite{strümpfer2012open}, as dissipation affects electron and exciton transfer~\cite{nitzan2024chemical,may2023charge,schlimgen2021quantum}, and it plays a major role in biological processes, e.g. photosynthesis~\cite{blankenship2021molecular}.

Various analytical and numerical approaches have been introduced, including quantum trajectories~\cite{dalibard1992wave,daley2014quantum}, Matrix Product Operator methods~\cite{verstraete2004matrix,zwolak2004mixed,verstraete2008matrix}, neural networks \cite{hartmann2019neural}, to cite but a few (see~\cite{weimer2021simulation} for a review).

In this work, we revisit an approach ---adopted for instance in Refs.~\cite{lange2017pumping,lange2018time,lenarvcivc2018perturbative,bouchoule2020effect,bouchoule2021breakdown,rossini2021strong,rosso2021one,reiter2021engineering,rosso2022one,riggio2023effects,maki2024loss,gerbino2024large,gerbino2024kinetics,rowlands2024quantum,zundel2024space,ulvcakar2024generalized}--- aimed specifically at simulating the {\it slow dynamics} of quantum many-body systems with {\it weak dissipation}. The main idea can be sketched as follows. Starting from the Markovian dynamics of an extended open quantum many-body system modeled by the Lindblad equation for the density matrix $\hat{\rho}$,
\begin{equation}
    \partial_t\Hat\rho=-i[\Hat H,\Hat\rho] +\gamma \sum_{\alpha} \left(\Hat L_{\alpha}\Hat\rho \Hat L^\dagger_{\alpha}-\dfrac{1}{2}\{\Hat L_{\alpha}^\dagger \Hat L_{\alpha},\Hat\rho\}\right) ,
 \label{eq:Lindblad}
\end{equation}
where $\hat{H}$ is the Hamiltonian governing the unitary part of the dynamics and the $\hat{L}_\alpha$'s are the dissipators, one focuses on the regime where the dissipation is weak, i.e. $\gamma \rightarrow 0$. In this regime of “adiabatic dissipation”, the system undergoes a separation of time scales, between the fast unitary dynamics and the slow dissipation. For a chaotic Hamiltonian $\hat{H}$ with no conserved quantities, the fast unitary dynamics of the extended system is expected to enforce quick local relaxation to a Gibbs ensemble, so that the density matrix of the system can be approximated as
\begin{equation}
\label{eq:adiabatic_assumption}
\hat{\rho}(t) \simeq \hat{\rho}_{\beta(t)} = \frac{1}{Z} e^{- \beta(t) \hat{H} } . 
\end{equation}
As usual, this approximation is valid in the sense that expectation values of local observables evaluated w.r.t $\hat{\rho}(t)$ and $\hat{\rho}_{\beta(t)}$ are nearly identical. In this case, the slow dynamics of the system can be effectively encoded in a single parameter: the slowly-varying inverse temperature $\beta(t)$. An evolution equation for that parameter is obtained by requiring that the mean energy, $\langle \hat{H} \rangle_\beta = {\rm tr}[\hat{\rho}_\beta  \hat{H}]$, evolves according to Eq. (\ref{eq:Lindblad}), leading to $\partial_t \langle \Hat H \rangle_{\beta (t)} = \gamma \, {\rm Re}  \left( \sum_{\alpha}  \langle [ \Hat L^\dagger_\alpha , \Hat H ] \Hat L_\alpha \rangle_{\beta(t)} \right) $. This is a closed evolution equation for $\beta(t)$, whose solution is clearly of the form $\beta(t) = f(\gamma t)$ for some function $f$. That function fully determines the dynamics of the system on time scales $t \sim O(1/\gamma)$. This result holds under the assumption that the approximation  (\ref{eq:adiabatic_assumption}) is valid.

The above logic generalizes straightforwardly to quantum many-body systems with more conservation laws, e.g. particle number conservation. In the series of recent works ~\cite{lange2017pumping,lange2018time,lenarvcivc2018perturbative,bouchoule2020effect,bouchoule2021breakdown,rossini2021strong,rosso2021one,reiter2021engineering,rosso2022one,riggio2023effects,Perfetto_PRL2023,Perfetto_PRE2023,gerbino2024kinetics,gerbino2024large,rowlands2024quantum,zundel2024space,ulvcakar2024generalized}, this approach has also been applied to quantum integrable systems, where the number of conservation laws scales extensively with system size. In that case, the approximate density matrix (\ref{eq:adiabatic_assumption}) is replaced by a Generalized Gibbs Ensemble (GGE)~\cite{rigol2007relaxation,rigol2008thermalization,vidmar2016generalized,essler2016quench}. This “time-dependent GGE” ($t$-GGE) approach to quantum many-body dynamics in this regime of {\it adiabatic dissipation} has been applied recently to characterize the steady state of weakly driven nearly integrable systems \cite{lange2017pumping,lange2018time,lenarvcivc2018perturbative}, to study the effects of atom losses in the Lieb-Liniger  gas~\cite{bouchoule2020effect,bouchoule2021breakdown,maki2024loss} as well as losses or loss/gain or reaction-diffusion models in 1D lattice gases \cite{rosso2021one,rossini2021strong,rosso2022one,riggio2023effects,gerbino2024kinetics,gerbino2024large,rowlands2024quantum,zundel2024space}, or to propose implementations of interesting GGEs on noisy quantum devices~\cite{reiter2021engineering,ulvcakar2024generalized}.

\begin{figure*}[t]
\centering
    \includegraphics[width=0.48\textwidth]{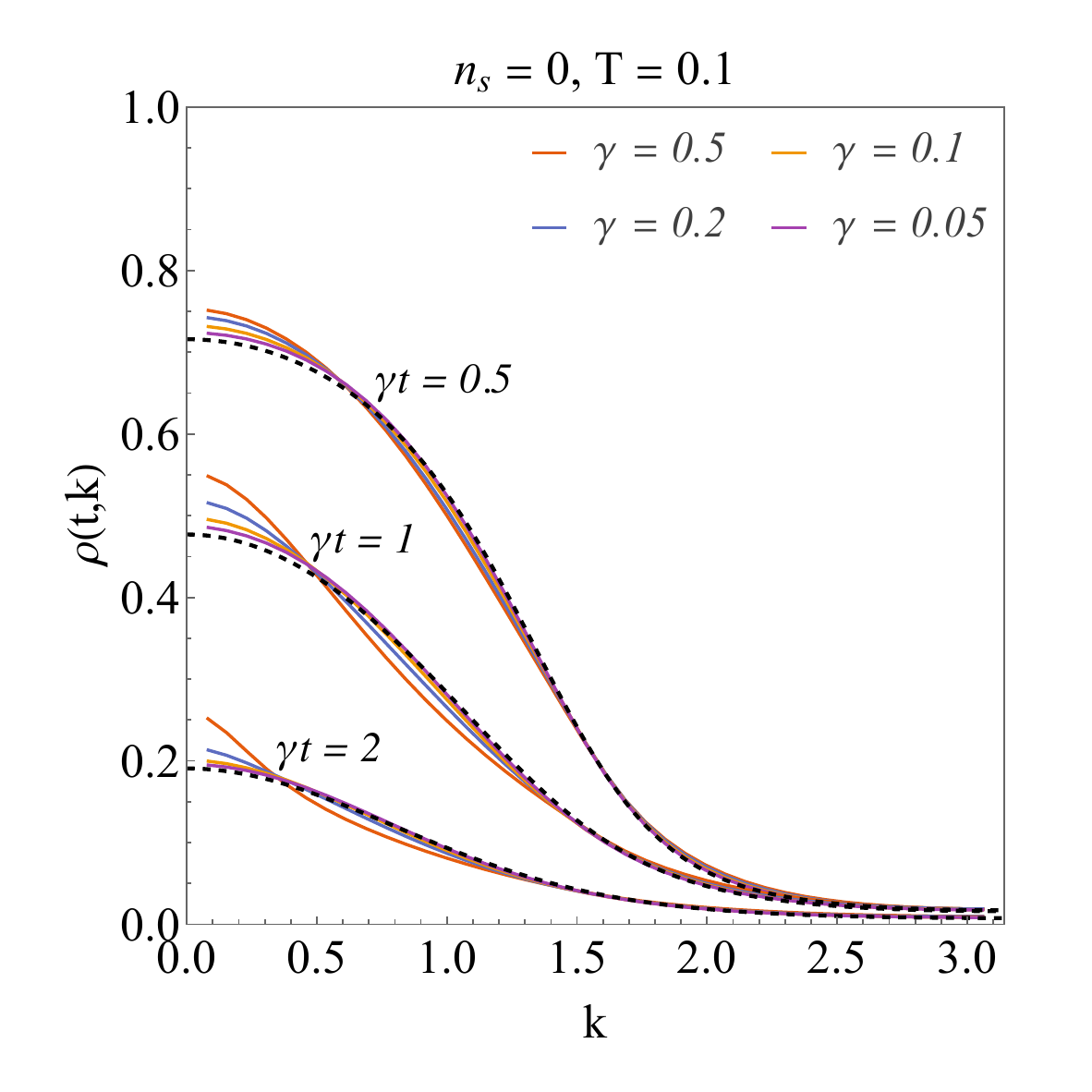}
    \includegraphics[width=0.48\textwidth]{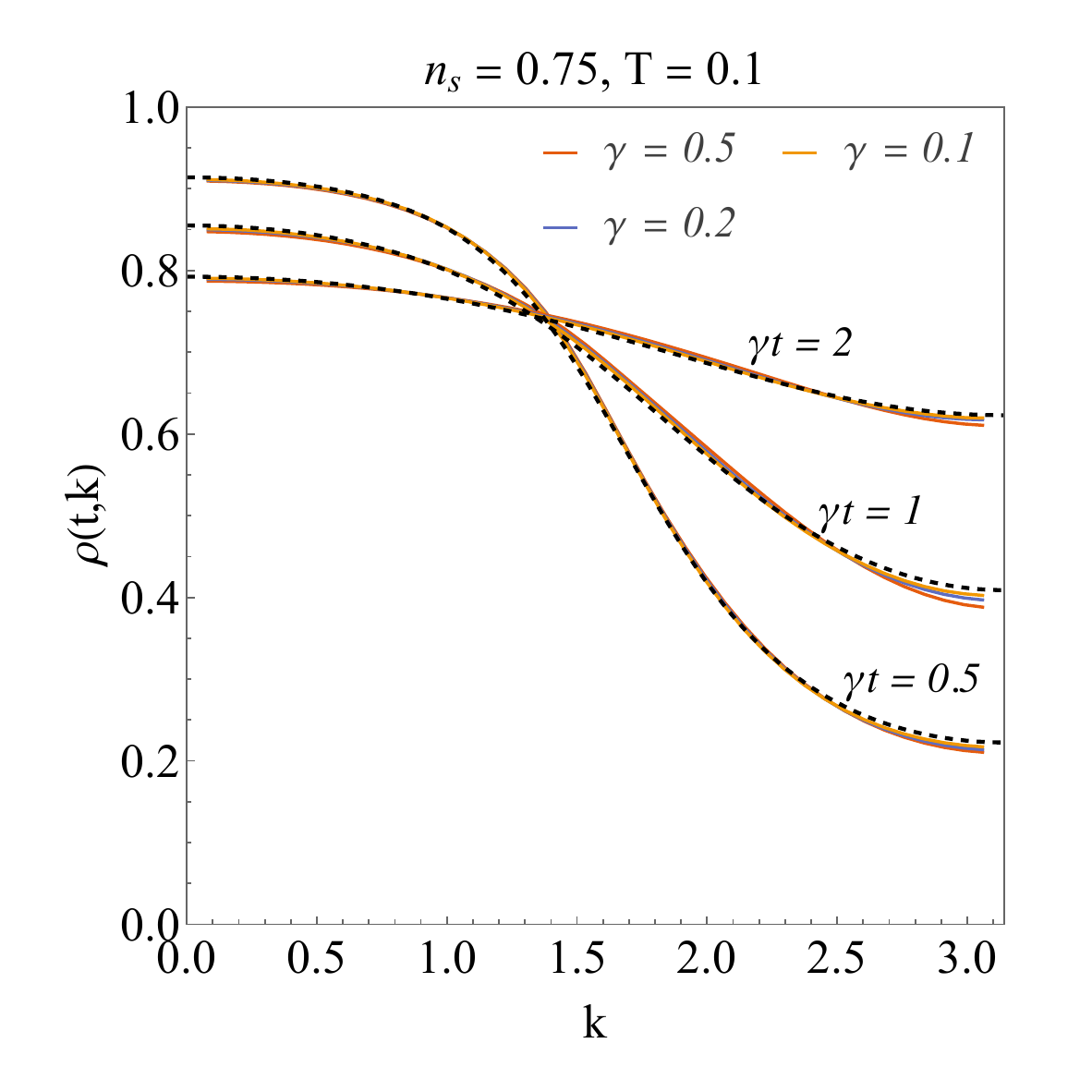}
    
\caption{Comparison of the rapidity distribution $\rho(t,k)=\tr[\hat{n}(k)\hat\rho(t)]$ for the exact numerical evolution (full-color lines) and the analytical prediction obtained with the $t$-GGE approximation (dashed black lines) at fixed $\gamma t\in \{0.5,1,2\}$ for decreasing rate $\gamma$, and two values of the stationary density $n_s=0$ (left), $n_s=0.75$ (right). We have fixed the energy unit to $J=1$. The initial state is prepared in the canonical ensemble at temperature $T=0.1$. We clearly see that the exact results at finite $\gamma$ go to the analytical curve predicted by the $t$-GGE assumption. For more details, see Section~\ref{sec:homogeneous}.
} 
\label{fig:summary_nk}
\end{figure*}

We stress that the validity of the $t$-GGE approximation, i.e. the analog of Eq.~(\ref{eq:adiabatic_assumption}) for infinitely many conservation laws, is not obvious. Indeed, one could be tempted to argue that, if there was a finite relaxation time $\tau$ such that after time $\tau$ the system locally relaxes to a GGE, then the approach should be valid as long as $\gamma \tau \ll 1$. The problem with that argument is that the local relaxation towards the GGE is algebraic rather than exponential~\cite{fagotti2013reduced,lux2014hydrodynamic,essler2023short}, so $\tau$ is infinite, and it is not obvious that $\gamma$ can ever be taken small enough. Moreover, several works have stressed that the $t$-GGE assumption can in principle be used in combination with Generalized Hydrodynamics~\cite{bertini2016transport,castro2016emergent} to describe the hydrodynamic evolution, in the presence of dissipation, from inhomogeneous initial states or in external trapping potentials~\cite{bouchoule2020effect,rosso2022one,riggio2023effects,maki2024loss}. In such inhomogeneous situations, it is even less clear that the $t$-GGE assumption remains accurate at all times and everywhere in the system. 

Therefore, the $t$-GGE approximation is worth putting to the test. To our knowledge, only a handful of works have done so and provided direct evidence for its validity and accuracy. Refs.~\cite{lange2017pumping,lange2018time,lenarvcivc2018perturbative,reiter2021engineering} provided direct checks against exact diagonalization results in homogeneous integrable spin chains that were, however, limited to small system sizes. Refs.~\cite{rossini2021strong,rosso2021one} checked the predictions of the $t$-GGE approach against quantum trajectory simulations in homogeneous quantum gases with atom losses; those checks were also performed for relatively small system sizes. Analytical arguments from Keldysh theory were recently developed for reaction-diffusion models in Refs.~\cite{gerbino2024large,gerbino2024kinetics}.

In this paper, our objective is twofold. First, we aim at providing numerical benchmarks and compelling evidence for the exactness of the $t$-GGE approach in the $\gamma \rightarrow 0$ limit, for unprecedently large system sizes and long times. Second, we aim to test the validity of the $t$-GGE approximation combined with Generalized Hydrodynamics in an inhomogeneous setting.

\subsection{Summary of the main results}

We compare the predictions of the $t$-GGE approach to numerically exact Matrix Product Operator (MPO) simulations of the dissipative dynamics. Such MPO simulations are typically limited by a “barrier” of operator entanglement at intermediate times~\cite{wellnitz2022rise,preisser2023comparing}, however here we select a model ---the hardcore boson gas subject to one-body bosonic losses and gain--- for which this barrier is moderate, allowing for numerically exact simulations of the microscopic dynamics for very long times. An additional advantage of the selected model is that the closed evolution equation derived from the $t$-GGE approach (i.e. the analog of the equation satisfied by $\beta(t)$ above) can be written in a fully explicit form and can even be solved analytically in some cases~\cite{bouchoule2020effect,riggio2023effects} (see Eq.~(\ref{eq:effective eq}) below). Thus, the derivation of quantitative predictions from the $t$-GGE approach is very transparent in this model, further facilitating the comparison with the numerically exact simulation.

In the hardcore boson gas, the GGE is fully parametrized by its {\it distribution of rapidities} (i.e. the distribution of momenta of the underlying Jordan-Wigner fermions, see Sec.~\ref{sec:tGGEreview} for details). Thus, to check the validity of the $t$-GGE approach in the homogeneous case, we compute the distributions of rapidities by MPO simulations for different dissipation strengths $\gamma$ and at different times $t$, and compare them with the expectation from the $t$-GGE approach. For definiteness, we focus on the evolution of a homogeneous gas in a flat box, initialized at thermal equilibrium. Our results are shown in Fig.~\ref{fig:summary_nk}: we clearly observe that the rapidity distributions collapse on a function of $\gamma t$ when $\gamma \rightarrow 0$ and $t \rightarrow \infty$, as expected from the $t$-GGE approach. Moreover, the limiting rapidity distribution is in perfect agreement with the prediction of the $t$-GGE approach given by Eq.~(\ref{eq:effective eq}). Importantly, GGEs for the hardcore boson gas are equivalent to Gaussian density matrices for the Jordan-Wigner fermions, so checking the validity of the $t$-GGE approach in that model is equivalent to checking that the density matrix remains approximately Gaussian. Therefore, we also study the accuracy of the approach beyond two-point function, introducing an appropriate estimate of non-Gaussianity for mixed states that can be efficiently evaluated as a tensor network, see Eq.~(\ref{eq:delta_rho_rho}). Our numerical results (Fig.~\ref{fig: purity}) show that the density matrix becomes increasingly Gaussian as $\gamma$ decreases.

Then, to check the validity of the $t$-GGE approach combined with Generalized Hydrodynamics (GHD), we focus on a paradigmatic situation of a spatially inhomogeneous setting: two half-systems initially prepared at different densities. We follow the density profile evolution under the joint effects of ballistic transport of the quasi-particles and dissipation due to loss and gain. We compare our MPO simulations to the prediction of the dissipative GHD equation (\ref{eq:evolution_transport}) for gain and loss, which we solve numerically. Again, we find excellent agreement for weak dissipation, see Fig.~\ref{fig:transport}.

\subsection{Organization of the paper}

The paper is organized as follows. The lattice hardcore boson model with one-body losses and gain is introduced in Section~\ref{sec:model}. We briefly review the $t$-GGE approach to that model in Section~\ref{sec:tGGEreview}; in particular, we present helpful analytical formulas adapted from Ref.~\cite{riggio2023effects} that are used later for the comparison with MPO simulations. The MPO method is reviewed in Section~\ref{sec: TN sol}. We present our numerical results in Sections \ref{sec:homogeneous} and \ref{sec:inhomogeneous}, for the homogeneous and inhomogeneous settings respectively. We conclude in Section~\ref{sec:conclusion}.

\section{The model and its stationary state}
\label{sec:model}

We consider a gas of hardcore bosons in a one-dimensional lattice, subject to gain and loss of bosons. The model corresponds to the infinite repulsion limit of the Bose-Hubbard model, commonly used to describe bosonic gases at low temperatures, see e.g. Refs.~\cite{ronzheimer2013expansion,vidmar2015dynamical}. Annihilation and creation of bosons at a site $j$ can be represented by Pauli $\hat\sigma_j^+,\,\hat\sigma_j^-$ operators, where $\hat\sigma^\pm_j = \frac12(\hat\sigma_j^x\pm i\hat\sigma_j^y)$. These operators automatically take care of the constraint that there is never more than one boson on site $j$ because $(\hat\sigma^\pm_j)^2=0$. At different sites $i\neq j, \,\,[\hat\sigma^+_i,\hat\sigma^-_j]=0$, and the local boson density is $\hat n_j=\hat\sigma_j^-\hat\sigma_j^+ =  \frac12(1 - \hat\sigma_{j}^{z})$. [Note that $\hat\sigma^{z} = Z = \left| 0 \right> \left< 0 \right| - \left| 1 \right> \left< 1 \right|$, so with this convention, a site occupied by a boson corresponds to a qubit in the state $\left| 1 \right>$.]
The unitary dynamics is governed by the nearest-neighbor hopping Hamiltonian
\be\label{eq:H}
\hat H = -\frac{J}{2} \sum_{j=1}^L \left( \hat\sigma_{j}^{+}  \hat\sigma_{j+1}^{-}
+\hat\sigma_{j}^{-}  \hat\sigma_{j+1}^{+} \right)\,,
\ee
which conserves the total number of bosons $\hat{N}=\sum_j \hat{n}_j$.
In addition, we assume that the gas is subject to incoherent single-particle loss and gain processes, resulting in a Markovian evolution  governed by the following Lindblad equation,
\begin{equation}
\begin{split}
    \partial_t\Hat\rho=-i[\Hat H,\Hat\rho]&+\gamma_L \sum_{j=1}^L\left(\Hat \sigma_{j}^+ \Hat\rho \Hat \sigma^-_{j}-\dfrac{1}{2}\{ \Hat \sigma^-_{j} \Hat \sigma^+_{j},\Hat\rho\}\right)\\
    &+\gamma_G\sum_{j=1}^L\left(\Hat  \sigma^-_{j} \Hat\rho \Hat \sigma^+_{j} -\dfrac{1}{2}\{\Hat \sigma^+_{j} \Hat \sigma^-_{j},\Hat\rho\}\right), \label{eq:Lindblad Gain Loss}
\end{split}
\end{equation}
where $\gamma_L,\gamma_G > 0$ set the loss and gain rates,
and the dissipator $\hat\sigma^\pm_j$ removes/adds a particle at site $j$. 
The jump operators do not conserve the total number of bosons $\hat{N}$, however, if at $t=0$ the density matrix is block-diagonal with blocks corresponding to eigenspaces of $\hat{N}$, then it remains block-diagonal at any later time $t>0$. This is due to a weak symmetry of the Lindbladian, as explained in detail in Appendix~\ref{A:GainFun}.
%

The Lindbladian is uniform, hence we expect the density matrix $\hat\rho$ to relax to a homogeneous stationary state at very long times. A natural candidate is the infinite-temperature state at fixed chemical potential $\mu$, 
\begin{equation}
    \label{eq:rho_infty}
    \hat{\rho}_{s} \, = \, \frac{1}{Z} e^{\mu \sum_j \hat n_j} = \prod_{j=1}^L \frac{e^{\mu \hat n_j}}{1+e^{\mu}} .
\end{equation}
Indeed, since the total number of bosons is conserved, it is clear that the commutator $-i [\hat{H}, \hat{\rho}_{s} ]$ in (\ref{eq:Lindblad Gain Loss}) vanishes for any $\mu$. Moreover, that state is annihilated by the sum of the onsite loss and gain dissipators in (\ref{eq:Lindblad Gain Loss}), for any fixed $j$, provided that the particle density at site $j$ is
${\rm tr } [\hat{\rho}_s \hat{n}_j] = n_s$ where 
\begin{equation}
    \label{eq:stationary}
     n_s  \equiv  \frac{\gamma_G}{\gamma_G + \gamma_L} .
\end{equation}
The density matrix (\ref{eq:rho_infty}) is therefore a stationary state for the Lindblad equation (\ref{eq:Lindblad Gain Loss}) if the chemical potential is $\mu=\log( \gamma_G/\gamma_L)$, and we see that $n_s$, as defined in Eq.~(\ref{eq:stationary}), is the stationary atom density at infinite times.

This model is a generalization of the one in
Ref.~\cite{riggio2023effects}, which considered only losses, and where the stationary state was the vacuum. Let us stress that, even if determining the stationary state $\hat\rho_s$ is an easy task, the Lindblad equation in Eq.~\eqref{eq:Lindblad Gain Loss} is not exactly solvable. This is in stark contrast with the purely fermionic models considered, for instance, in Refs.~\cite{alba2021spreading,carollo2022dissipative,alba2023logarithmic}, where the Lindblad dissipators $\hat{\sigma}^\pm_j$ are replaced by fermionic creation/annihilation operators. In that case, the state $\hat{\rho}(t)$ remains exactly Gaussian at any time and can be computed with free fermion techniques. Here, in our model with {\it bosonic} loss and gain (\ref{eq:Lindblad Gain Loss}), the dynamics cannot be solved exactly. 
Describing the dynamics at intermediate times, when the system has not yet relaxed to the steady state, is a challenging problem that requires relying on approximate schemes. This is where the $t$-GGE approach comes in.

\section{$\mathbf{t}-$GGE approach}
\label{sec:tGGEreview}

In this section, we apply the $t$-GGE approach to the hardcore boson model, including gain and loss. We follow the discussion of Ref.~\cite{riggio2023effects}, which treats the case of losses only (i.e. $\gamma_G = 0$) and arrive at new formulas for the case including gain of atoms. As sketched in the introduction, the idea behind the approach is to adopt an adiabatic point of view and focus on the slow evolution of the conserved quantities that commute with the Hamiltonian~\eqref{eq:H} (but not with the dissipators). The conserved quantities specify the GGE towards which subsystems relax under unitary dynamics. The dissipative processes break the conservation laws and, therefore, bring the system away from that GGE. We then make the assumption of {\it adiabatic dissipation}: for slow dissipative processes, we assume that the system has enough time to equilibrate between consecutive dissipative processes. We require a separation between the fast time-scale of microscopic interactions $\sim1/J$ and the dissipative time-scale $\sim 1/(\gamma_G+\gamma_L) $:
\begin{equation}
    \gamma \equiv \gamma_{\rm G} + \gamma_{\rm L}  \, \ll \, J \, .
\end{equation}
In that regime, dissipative processes induce a slow evolution of the parameters of the GGE. We now turn to the main points of that approach for our model of bosonic gain and loss, starting first with homogeneous initial states and then turning to inhomogeneous states. More technical details are discussed in App.~\ref{A:GainFun}.

\subsection{Homogeneous dynamics}

Consider the gas of hardcore bosons initially prepared in a GGE state. All conserved charges can be found by diagonalizing the Hamiltonian in Eq.~(\ref{eq:H}) in terms of Jordan-Wigner fermions
\be
\hat c_j = \prod_{i<j} \hat \sigma^{z}_{i} \; \hat\sigma^{+}_{j}\,,
\ee
whose Fourier modes $\hat c_k = \frac{1}{\sqrt{L}} \sum_{j=1}^L e^{-i k j}\hat c_j$ diagonalize the Hamiltonian (\ref{eq:H}): $\hat H = \sum_{k} \epsilon(k) \hat n(k)$, where $\epsilon(k)=- J \cos(k)$ and
\begin{equation}
    \label{eq:nk}
\hat n(k)=\hat c^\dagger(k) \hat c(k) . 
\end{equation}
 The mode occupations $\hat n(k)$ are the
 conserved charges that define the GGE
\be\label{eq:rho_gge}
\hat\rho_{\rm GGE} = \frac{1}{Z} \exp \biggl( -\sum_{k} \nu_{k}\, \hat n(k) \biggr),
\ee
where $Z = {\rm tr}[\exp \left( -\sum_{k} \nu_{k} \hat \rho_{k} \right)] = \prod_k (1+e^{-\nu_k})$. We stress that, here, because the Hamiltonian is the one of non-interacting fermions, the GGE is always a Gaussian density matrix. 

In the $t$-GGE approach, the effective dynamics is restricted to the manifold of Gaussian states. This is only an approximation: although the Hamiltonian in the fermionic formulation is quadratic, the Jordan-Wigner strings carried by the Lindblad dissipators $\hat{\sigma}^\pm_j$ do not preserve Gaussianity exactly.
Assuming that the density matrix is a GGE at any time, the evolution of that GGE is encoded in the time dependence of its Lagrange multipliers $\nu_k(t)$ or, equivalently, in its {\it rapidity distribution}
\begin{equation}
\label{eq: rapidity distribution}
\rho(t,k) = {\rm tr}\bigl[\hat n(k) \hat\rho_{\rm GGE}(t)\bigr] = \frac{1}{1+e^{\nu_k(t)}}\,.
\end{equation}
Plugging the Lindblad equation~(\ref{eq:Lindblad Gain Loss}) into $\partial_t \rho(t,k) = {\rm tr}[\hat n(k)\partial_t {\hat\rho}]$, one can derive a closed non-linear time-evolution equation that must be satisfied by the rapidity distribution $\rho(t,k)$,
\begin{equation}
\label{eq:self_consistent}
    \partial_t\rho(t,k)=-\gamma_L \mathcal{F}_L[\rho](k)-\gamma_G \mathcal{F}_G[\rho](k),
\end{equation}
where the loss and gain functionals are defined as
\begin{align}
    \label{eq:F_L}
    \mathcal{F}_{L}[\rho](k)&=\sum_{j=1}^L\text{Re\,tr}\langle \Hat \sigma_{j}^+ [\Hat \sigma^-_{j},\Hat{n}(k)]\rangle_{\hat\rho_{\rm GGE}} \\ 
    \label{eq:F_G}
    \mathcal{F}_{G}[\rho](k)&=\sum_{j=1}^L\text{Re\,tr}\langle \Hat \sigma^-_{j}[\Hat \sigma_{j}^+,\Hat {n}(k)] \rangle_{\hat\rho_{\rm GGE}} \,.
\end{align}
Note that the expectation values in the r.h.s of (\ref{eq:F_L})-(\ref{eq:F_G}) are evaluated in the GGE whose Lagrange multipliers have been fixed by $\rho(t,k)$. This means that ${\rm tr}[\hat c^\dagger(k) \hat c(k') \hat\rho(t)] = \rho(t,k)\,\delta_{k,k'}$, and the crucial simplification is that in Eqs.~(\ref{eq:F_L})-\eqref{eq:F_G} the expectation values can be analytically computed using Wick's theorem from the two-point function above. Some care in the calculation is needed because of the Jordan-Wigner strings, as already pointed out in Refs.~\cite{bouchoule2020effect,riggio2023effects}. The issue is discussed in Appendix~\ref{A:GainFun}, where we provide the complete derivation. 
%
%
%
%
    %
In the thermodynamic limit, one finds that the dissipation functionals are given in terms of the Hilbert transform, which for $2\pi$-periodic functions takes the form
\begin{equation}
    \mathcal{H}[f](x)=\dfrac{1}{2\pi}\,P.V.\!\!\int^\pi_{-\pi} dy~\dfrac{f(y)}{\tan(\frac{x-y}{2})}\,,
\end{equation}
where $P.V.$ denotes the Cauchy principal value of the integral. The dissipation functionals can be expressed in closed form as
\begin{align}
    \label{eq:F_gauss L}
    \mathcal{F}_{L}[\rho]&=\rho-\rho^2-\mathcal{H}^2\left[\rho\right] +n^2 +2n\mathcal{H}'\left[\rho\right]
    \\
    \label{eq:F_gauss G}
    \mathcal{F}_{G}[\rho]&=-\rho+\rho^2-\mathcal{H}^2\left[\rho\right]-\left(1-n\right)^2\\
    &\hspace{12pt}+2(1-n)\mathcal{H}'\left[\rho\right]\,,\nonumber
\end{align}
where $n(t) =\langle \Hat{N}\rangle/L$ denotes the mean boson density
\begin{equation}
    n(t)=n_s+\left(n_0-n_s\right)e^{-\gamma t} ,
\end{equation}
for an initial density $n_0$ and a steady-state density $n_s$. We have left intended the $t,k$ dependencies to avoid cluttering in the notation: as $\rho=\rho(t,k)$, the functionals and the Hilbert transform are calculated over $\rho$ at a given time $t$ for all $k$ like in $\mathcal{H}[\rho(t,\cdot)](k)$.
Plugging the expressions above into Eq.~\eqref{eq:self_consistent} provides the effective evolution equation describing our system, which we rewrite 
\begin{equation}
\label{eq:effective eq}
    \partial_{\gamma t}\rho = - \mathcal{F}[\rho]\,,
\end{equation}
having defined 
\begin{equation}
 \mathcal{F}[\rho] \equiv (1-n_s) \mathcal{F}_L[\rho] + n_s\mathcal{F}_G[\rho] \,.
\end{equation}
The effective dynamics of the system therefore depends on $t,~\gamma_L,~\,\gamma_G, \text{ and }J$ only through the rescaled time $\tau = \gamma t$ and the stationary density $n_s = \gamma_G / (\gamma_G + \gamma_L)$. In particular, it does not depend on the microscopic time-scale $J$ at all. 
%
It turns out that the evolution equation~\eqref{eq:effective eq} can be solved analytically by generalizing the techniques of Refs.~\cite{bouchoule2020effect,riggio2023effects} to the case with both loss and gain. The general solution reads
\begin{multline}
    \label{eq:analytic_solution}
    \rho(\tau,z)=n(\tau)\\+\text{Re}\left[\dfrac{(n_0-i\mathcal{H}[\rho_0](z))f(t)}{1+(2n_s-1)(n_0-i\mathcal{H}[\rho_0](z))\int_0^\tau dy~f(y/\gamma)}\right],
\end{multline}
where $z=z(\tau,k)\equiv k+g(t)$, with $g(t)=2i\left(e^{-\tau}-1\right)(n_0-n_s)(2n_s-1)+4in_s(1-n_s)\tau$, while $f(t)=e^{-ig(t)}e^{-\tau}$, and $\rho_0(k)$ is the initial density profile. The derivation of this exact solution is provided in Appendix \ref{A:MastEq SignalF}. Notably, this form of the rapidity distribution holds if dephasing and incoherent hopping are added on top of the one-body gains and losses, only $f(t)$ has to be modified to consider the new dissipative effects. It is also worth stressing that this expression allows fast computations for an arbitrary choice of parameter $n_s$ and $\tau$. Indeed, as outlined in Appendix \ref{A:Numerical ev of H}, calculating a Hilbert transform numerically presents no technical challenges, which is also true for the $f$ integral. Additionally, in the scenario of balanced gains and losses $\gamma_L=\gamma_G$, this equation can be greatly simplified, resulting in a density profile $\rho(t,k)=\text{Re}[i\mathcal{H}[\rho_0](z(\gamma t,k))]$.

\subsection{Inhomogeneous dynamics: dissipative GHD}
\label{sec:tGGE-GHD}

Typical experimental conditions for quantum gases break translational invariance, e.g. because of confining potentials or inhomogeneous initial conditions. In such inhomogeneous settings, generic many-body systems are expected to evolve according to an emergent hydrodynamic theory at large scales. Here, for hardcore bosons, which constitute an integrable model, the right setup is Generalized Hydrodynamics (GHD) \cite{bertini2016transport,castro2016emergent}. The main idea is again the separation of time scales between the fast underlying microscopic dynamics and the slow macroscopic evolution of conserved charge densities and currents, such that the gas locally relaxes to a GGE parameterized by the local rapidity distribution $\rho(t,x,k)$. For hardcore bosons without dissipation, the time-evolution of the rapidity distribution is
\begin{equation}
\label{eq:GHD}
    \partial_t \rho(t,x,k)+v(k)    \partial_x\rho(t,x,k)=0\,,
\end{equation}
where the second term is the current of quasi-particles with momentum $k$, $j(t,x,k)=v(k) \rho(t,x,k)$, and
\begin{equation} 
    v(k)=\partial_k \epsilon(k) = J \sin k
\end{equation}
is their group velocity. 
At any finite time $t$, this description is only an approximation because, in reality, integrable models do not relax locally at any finite scale~\cite{fagotti2013reduced,lux2014hydrodynamic,essler2023short}. Still, this description becomes exact in the “Euler scaling limit.” This means
the following. Eq.~\eqref{eq:GHD} is invariant under the scale transformation $x\to \lambda x,\,t\to\lambda t$, if the initial condition is also rescaled accordingly, i.e. $\rho(0,x,k)  = f( \lambda x,k )$ for some function $f$. Then the dynamics of the finite system will collapse on the GHD solution only in the limit of infinite system sizes and infinite times, i.e. $\lambda \rightarrow 0$ and $x,t \rightarrow \infty$ so that $\lambda x$ and $\lambda t$ are finite.

Now let us include the effects of dissipation into  Eq.~(\ref{eq:GHD}). The separation of time scales required by GHD is the same as the one we have used in our approximation of adiabatic dissipation, so we expect the hydrodynamic picture to remain valid under the same assumption. Therefore, the slow dynamics of the system should be encoded in a space- and time-dependent GGE subject to gain and loss. The effect of the latter is given by the dissipation functionals previously introduced in Eqs.~\eqref{eq:F_L},~\eqref{eq:F_G}. The GHD equation is then modified to 
\begin{equation}
\label{eq:evolution_transport}
\partial_t \rho +v \,\partial_x\rho =-\gamma \mathcal{F}[\rho]\,,
\end{equation}
where the functional is now evaluated at every point $x$ and time $t$ and every rapidity $k$ as $\mathcal{F}[\rho(t,x,\cdot)](k)$. In this inhomogeneous setting, dissipation is locally homogeneous within a fluid cell, the only difference compared to the previous Section is that the term $\mathcal{F}[\rho]$ now depends on position through the spatial dependence of the rapidity distribution $\rho(t,x,\cdot)$.

In contrast with the homogeneous setting, for inhomogeneous initial states, the slow dissipative dynamics is not ruled solely by the rescaled time $\gamma t$ because the $\sim 1/J$ timescale now enters through the quasi-particle velocity $v(k)$. We see that Eq.~\eqref{eq:evolution_transport} is now invariant under the scale transformation $x\to\lambda x,\,t\to\lambda t,\,\gamma\to\gamma/\lambda$, if the initial condition scales as $\rho(0,x,k) = f(\lambda x,k)$. For inhomogeneous initial states, we thus expect the “dissipative GHD” description to become exact in the combination of Euler scaling limit and slow dissipation limit, i.e. when we send $x \rightarrow \infty$, $t \rightarrow \infty$, $\gamma \rightarrow 0$, keeping $x/t$ and $\gamma t$ fixed. One of the main goals of this paper is to check that this is indeed what happens. In Section~\ref{sec:inhomogeneous}, we will provide evidence for the validity of the “dissipative GHD” approach in that limit (see Fig.~\ref{fig:transport}).


\section{Matrix Product Operator simulation \label{sec: TN sol}}
\begin{figure}[t]
\centering
    \includegraphics[width=0.4\textwidth]{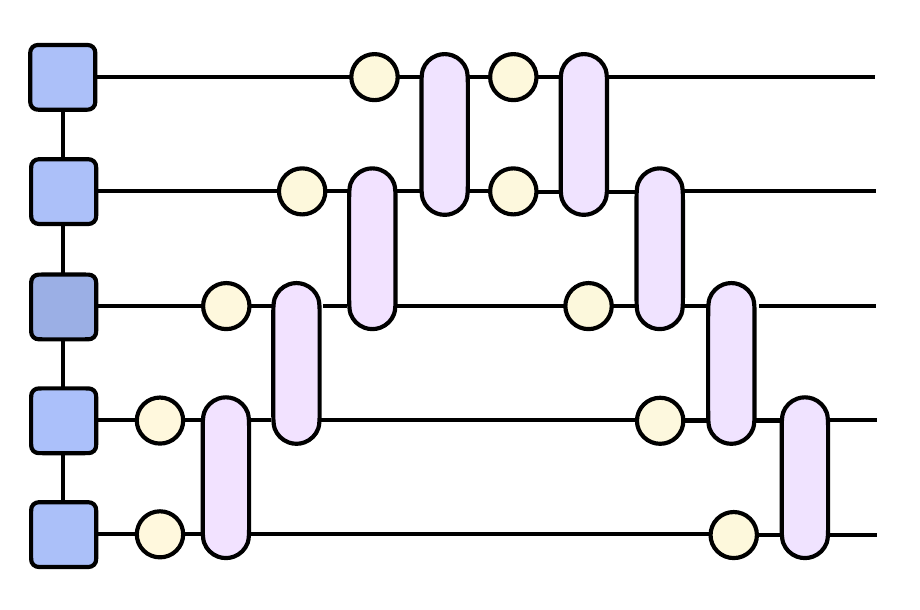}
\caption{Representation of the tensor network that implements a single time-step of the Lindblad density matrix evolution with single-particle dissipation. The yellow circles represent the action of the single-site dissipators, while the purple elements are the trotterized two-site unitary terms.} 
\label{fig: TN}
\end{figure}

A Lindblad evolution with local jump operators can be efficiently integrated as a tensor network with a modified Time-Evolving Block Decimation algorithm~\cite{vidal2004TEBD}. The update rule can be found by regarding $\hat\rho$ as a vector and finding the correct representation of the Lindbladian on the vector space to which the density matrix belongs. Considering the Pauli representation of hardcore bosons, we denote $|\hat\rho \rrangle$, the vectorized representation of $\hat\rho$. We take as local basis the normalized Pauli matrices $\tau_j^\mu\in \frac{1}{\sqrt{2}} \{\mathbb{I},\sigma_j^x,\sigma_j^y,\sigma_j^z\}$, then we expand the density matrix as a Matrix Product State (MPS)
\be
    |\hat\rho\rrangle = \sum_{\mu_1...\mu_L}\,R_{\mu_1}... R_{\mu_L}\, |\tau_1^{\mu_1}...\,\tau_L^{\mu_L}\rrangle\,.
\ee
 We implement the system with open boundary conditions. Even though the theoretical equations are derived for periodic boundaries, the difference is expected to vanish in the thermodynamic limit.
Separating the Hamiltonian as $\hat{H}=\sum_j \hat{H}_{j,j+1}$ and the dissipator into single site terms, the Lindblad generator can be represented as 
\be
    \mathscr{L} = \sum_{j=1}^{L-1} \mathscr{H}_{j,j+1} + \sum_{j=1}^L \mathscr{D}_j\,;
\ee
{$\mathscr{H}_{j,j+1}\!\sim\!-i[\hat{H}_{j,j+1},\boldsymbol{\cdot}\,]$ and $\mathscr{D}_j\! \sim\! \gamma_L \bigl(\hat{\sigma}^+_j\boldsymbol{\cdot}\hat{\sigma}_j^- -\frac12\{\hat{\sigma}_j^- \hat{\sigma}_j^+,\boldsymbol{\cdot}\}\bigr)$} 
$+\,\gamma_G \bigl(\hat{\sigma}^-_j\boldsymbol{\cdot}\hat{\sigma}_j^+ -\frac12\{\hat{\sigma}_j^+ \hat{\sigma}^-_j,\boldsymbol{\cdot}\,\}\bigr)$ are the linear representations of the super-operators on the right acting on the $|\hat\rho\rrangle$ space. The evolution of the density matrix is then
\be
|  \hat\rho_t\rrangle=e^{t\mathscr{L}}|\hat\rho_0\rrangle\,.
\ee
To find the update rule for the local tensors $R_{\mu_j}$, we rely on the Trotter expansion described in \cite{White2004_tDMRG}. If $\mathscr{U}_{j,j+1}=e^{dt\mathscr{H}_{j,j+1}}$, for a small step $dt$ we find
\begin{equation}
\begin{split}
\label{eq:trotter}
e^{dt \mathscr{L}}\approx 
e^{(dt/2)\mathscr{D}_1}\biggl(\prod_{j=1}^{L-1} e^{(dt/2)\mathscr{D}_{j+1}}\mathscr{U}_{j,j+1}\biggr) \cdot\\
 \cdot \,e^{(dt/2)\mathscr{D}_L}\biggl(\prod_{j=1}^{L-1} e^{(dt/2)\mathscr{D}_{L-j}}\mathscr{U}_{L-j,L-j+1}\biggr)
\end{split}
\end{equation}
%
The decomposition above leads to the tensor network represented in Fig.~\ref{fig: TN} for a single time step. The action of the super-operators above can be found by applying it to the Pauli strings, and rewriting the result back into the Pauli basis. Each dissipative term acts on a single site as 
\begin{equation*}
e^{(dt/2)\mathscr{D}_j}|\tau_j^\nu\rrangle = \sum_\mu d_{\mu\nu} |\tau_j^\mu\rrangle \,,\,\,\, d_{\mu\nu} = \llangle \tau_j^\mu | e^{(dt/2)\mathscr{D}_j} | \tau_j^\nu \rrangle\, 
\end{equation*}
because the $\tau$ matrices are orthonormal with respect to the Hilbert-Schmidt product $\llangle A | B \rrangle = {\rm tr}(A^\dagger B)$. Then
\begin{gather}
e^{(dt/2)\mathscr{D}_j}|\rho\rrangle = \sum_{\mu_1...\mu_L}\,R_{\mu_1}... R'_{\mu_j}...R_{\mu_L}\, |\tau_1^{\mu_1}...\,\tau_L^{\mu_L}\rrangle\,,\\
R_{\mu_j}'^{\,\alpha_j\alpha_{j+1}} = \sum_{\nu_j} d_{\mu_j\nu_j}R_{\nu_j}^{\alpha_j\alpha_{j+1}}\,.
\end{gather}
All dissipative terms are one-site and cannot increase the bond dimension, unlike the unitary term, which is two-site. In a similar way, $\mathscr{U}_{j,j+1}$ transforms $R_{\mu_j}R_{\mu_{j+1}}$ into
\be
     B_{\mu_j \mu_{j+1}} = \sum_{\nu_j,\nu_{j+1}} u_{\mu_j\mu_{j+1},\nu_j\nu_{j+1}} R_{\nu_j}R_{\nu_{j+1}}\,,
\ee
where $u_{\mu_j\mu_{j+1},\nu_j\nu_{j+1}} = \llangle \tau_j^{\mu_j}\tau_{j+1}^{\mu_{j+1}} | \mathscr{U}_{j,j+1} | \tau_j^{\nu_j}\tau_{j+1}^{\nu_{j+1}} \rrangle$ and the updated local tensors $R'$ come from the singular value decomposition (with a possible truncation of the bond dimension $\chi$)
\be
\begin{split}
B_{\mu_j \mu_{j+1}}^{\alpha_j\alpha_{j+2}}& = \sum_{\alpha_{j+1}} A^{\alpha_j\alpha_{j+1}}_{\mu_j}(\Lambda^{\alpha_{j+1}\alpha_{j+1}}B^{\alpha_{j+1}\alpha_{j+2}}_{\mu_{j+1}})  \,\\
&\to \sum_{\alpha_{j+1}=1}^\chi R'^{\,\alpha_j\alpha_{j+1}}_{\mu_j}R'^{\,\alpha_{j+1}\alpha_{j+2}}_{\mu_{j+1}}\,.
\end{split}
\ee
After the time evolution, the numerically exact occupation of the fermionic modes $\rho_{{\rm exact}}(t,k) = \text{tr}[ \hat{n}(k) \hat{\rho}(t)]$ is calculated from the contraction $\llangle \hat{n}(k) | \hat{\rho}(t)\rrangle$, where fermionic operators in this representation carry a Jordan-Wigner string.

Expanding the density matrix as an MPS over an operator basis is equivalent to representing it as an MPO. The drawback of this approach to time evolution is that the positivity of $\hat\rho$ is not guaranteed, because the Lindblad evolution preserves it but not the truncation of the bond dimension $\chi$. This issue could be circumvented by a slightly more expensive representation of $\hat\rho$ in its locally purified form \cite{werner2016lpdo}, but since negative eigenvalues are of the order of the truncation error in $\chi$ we choose to keep the MPO form. The small negative eigenvalues do not compromise the validity of the representation, since such dissipative processes generate states with low entanglement and the convergence with respect to the bond dimension is immediate. 
\begin{figure*}[t]
\includegraphics[width=\textwidth]{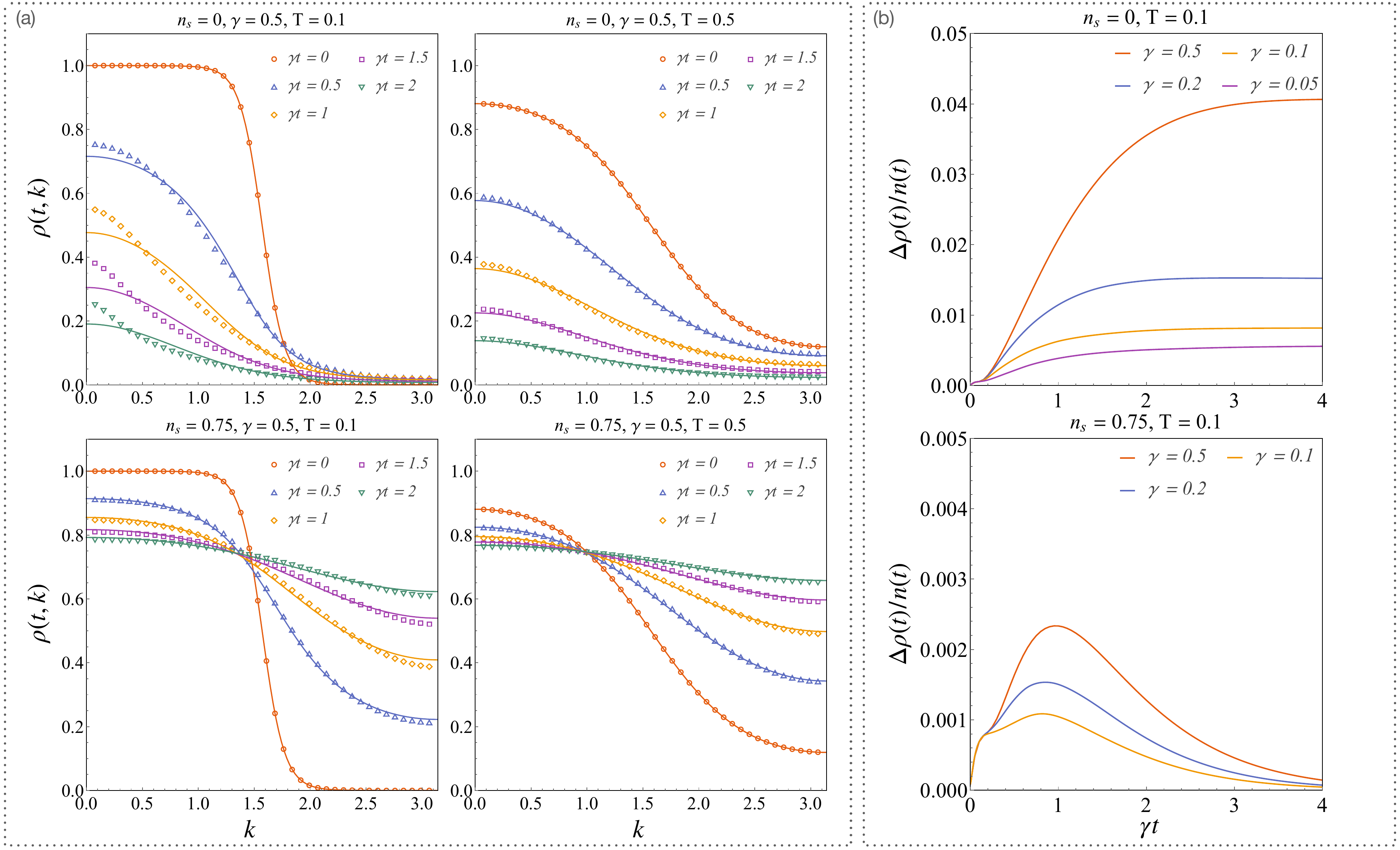}
\caption{(a) Comparison of the rapidity distribution $\rho(t,k)$, for the exact numerical evolution (markers) and the analytic approximation (curve) for different temperatures and stationary densities.
(b) Relative integrated error of the analytic solution compared to the exact numerical data. The parameters are the same as those in Fig.~\ref{fig:summary_nk}. See main text for details.} 
\label{fig:nk_T}\label{fig:relative_error}
\end{figure*}

\section{Accuracy of $\mathbf{t}-$GGE: homogeneous time evolution}
\label{sec:homogeneous}
We now turn to our main results: the numerical check of the validity of the $t$-GGE description in the hardcore boson gas with gains and losses. We prepare the system in a thermal initial state, $\hat\rho=\frac{1}{Z}e^{-\hat H/T}$, with the Hamiltonian \eqref{eq:H}. This is a Gaussian state for the Jordan-Wigner fermions. We stress once again that our 
Lindblad dissipators do not preserve Gaussianity, so under the Lindblad evolution the density matrix does not remain exactly Gaussian. 
The assumption that the density matrix is Gaussian is expected to be valid for large system sizes when $\gamma \ll 1$. In the following, we investigate the validity of that approximation numerically for finite $\gamma$.

\subsection{Results for rapidity distributions}

Here we compare the time evolution of the rapidity distribution $\rho(t,k)$ obtained by the $t$-GGE approach with the exact occupations of the diagonal modes $\rho_{{\rm exact}}(t,k)={\rm tr}[\hat n_k \hat \rho(t)]$ found by integrating the Lindblad Eq.~\eqref{eq:Lindblad Gain Loss} with tensor networks, as described above. The initial thermal state is prepared by imaginary time evolution, starting from an identity Pauli MPS. In this Section we work with a chain of $L = 40$ sites, and we have checked that this system size is large enough so that all results become size-independent. This is because in all our simulations, the correlation length in the system remains much smaller than $L$.

Some representative results have already been shown in Fig.~\ref{fig:summary_nk} where it is clearly visible that the discrepancy between the exact numerical data and the analytic approximation disappears in the limit of slow dissipation. Let us mention that the exact dynamics, in terms of the rescaled time $\gamma t$, still exhibits a residual dependence on the finite value of $\gamma$; however, it converges towards the $t$-GGE approximation in Eq.~\eqref{eq:effective eq} for any fixed $\gamma t$ in the limit $\gamma\to 0$.
Notably, the precision is improved at a finite stationary density, when there is a balance between loss and gain events. 
The temperature also plays a role. For sufficiently large temperature $T$, the adiabatic dissipation description is almost exact, indicating that the time scale associated with the microscopic dynamics is faster and the underlying bosons can locally relax to a GGE (see Fig.~\ref{fig:nk_T}).
Notice that the errors are picked up in the initial moments of the evolution and do not grow unbounded. We can see this explicitly by computing the total error at a fixed time
\begin{equation}
    \Delta\rho(t) = \int_{-\pi}^{+\pi}\frac{dk}{2\pi} \,\left\lvert \rho(t,k)- \rho_{\rm exact}(t,k) \right\rvert\,.
\end{equation}
In Fig.~\ref{fig:relative_error} we plot the relative error $\Delta\rho(t)/n(t)$, normalized by the total density $n(t)=\int_{-\pi}^{+\pi} dk/2\pi \rho(t,k)$, because when only losses are present $\Delta\rho(t)$ would trivially vanish as the state is depleted. Normalizing it shows that the accumulated discrepancy is lost along the evolution as the exact steady state is correctly captured by Eq.~\eqref{eq:effective eq}. Correspondingly, the relative error at a finite stationary density disappears for long times. 

Notably, the accuracy of the $t$-GGE approximation increases significantly for non-zero $n_s$ (in the case of gain and loss). Specifically, the accuracy is maximal when gain and loss are balanced ($n_s=1/2$) and decreases while considering only gain or loss ($n_s=0$ and $n_s=1$, respectively). We will discuss this behavior in more detail in Appendix \ref{A:Precision}.
\begin{figure*}[t]
\centering
  \includegraphics[width=0.9\textwidth]{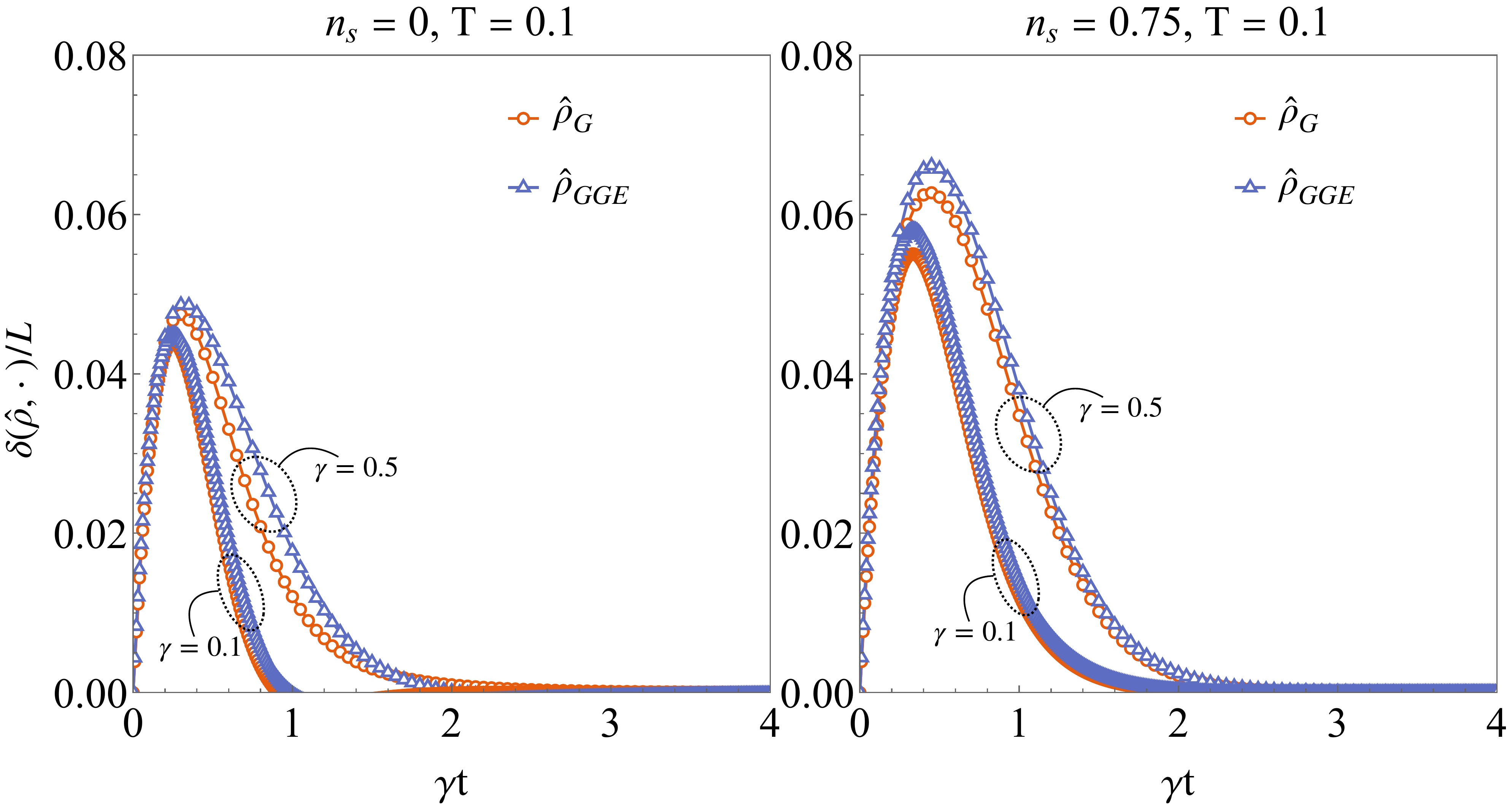}
\caption{R\'enyi-2 relative entropy defined in Eq.~(\ref{eq:delta_rho_rho}), normalized by the system length $L$, between the exact state $\hat{\rho}(t)$ and the optimal Gaussian approximation $\delta(\hat\rho,\hat\rho_G)$, and to the Gaussian state obtained from the analytical solution $\delta(\hat\rho,\hat\rho_{{\rm GGE}})$, for two different values of the total dissipation rate $\gamma = 0.5,\, 0.1$.} 
\label{fig: purity}
\end{figure*}

\subsection{A quantifier of non-Gaussianity}
In the previous Section, we have probed the rapidity distribution, which is a two-point function of Jordan-Wigner fermion creation/annihilation operators. Since bosonic dissipation breaks the Gaussianity of the model, higher correlation functions will not be given exactly by Wick's theorem, and the small error might proliferate in more complicated correlations. A compact way of checking the accuracy of the $t$-GGE description beyond single-particle observables is to evaluate the total non-Gaussianity of the system \cite{lumia2024, Genoni2008, Genoni2010}. 
%

The rapidity distribution obtained from the evolution equation~\eqref{eq:effective eq} parametrizes the Gaussian state $\hat\rho_{{\rm GGE}}$, see Eq.~\eqref{eq: rapidity distribution}. Knowing the solution of Eq.~\eqref{eq:effective eq}, one can the evaluate how close that Gaussian state is to the {\it exact} state $\hat\rho (t)$. The description above is a self-consistent Gaussian approximation of the full dynamics. Still, nothing guarantees that it is the \emph{optimal} Gaussian approximation of the exact state $\hat{\rho}(t)$. 
In fact it is clear that $\hat\rho_{{\rm GGE}}$ is not optimal; otherwise, we would have found precisely $\rho=\rho_{\rm exact}(t,k)$, and the difference would only be visible in higher correlations. 
However, we expect the largest component of the error to be brought by non-Gaussianity since the relative error stops increasing after the state has undergone enough dissipation to get close to being Gaussian again.

To quantify the non-Gaussianity contained in a state, let us consider the quantum relative entropy 
$S(\hat\rho||\hat\sigma) = {\rm tr}\bigl[\hat\rho(\log\hat\rho-\log\hat\sigma)\bigl]$, which is a common measure of distance between two states, even if it is not symmetric, unlike a true  metric 
\cite{vedral2002relativeentropy}.
We define the non-Gaussianity of $\hat\rho$ as the minimum relative entropy between $\hat\rho$ and the manifold of Gaussian states. If we vary $\hat\sigma$ over the Gaussian manifold, the minimization problem is solved by $\hat\sigma=\hat\rho_G$ where $\hat\rho_G$ the Gaussian partner of $\hat\rho$ built from the same two-point functions \cite{Marian2013}. Then the non-Gaussianity of $\hat\rho$ is measured by
\begin{eqnarray}
    \label{eq:quantity_not_good_for_calculation}
\nonumber S(\hat\rho||\hat\rho_G) &=& - {\rm tr}\bigl[\hat\rho \log\hat\rho_G \bigl] + {\rm tr}\bigl[\hat\rho \log\hat\rho\bigl] \\
\nonumber &=& - {\rm tr}\bigl[\hat\rho_G \log\hat\rho_G \bigl] + {\rm tr}\bigl[\hat\rho \log\hat\rho\bigl] \\ 
&= & S(\hat\rho_G) - S(\hat\rho)\,,
\end{eqnarray}
where $S(\hat\rho)=\tr [\hat\rho\log\hat\rho]$ is the Von Neumann entropy. To go from the first to  the second line, we have used two facts: (i) $\log\hat\rho_G$ is quadratic in the creation/annihilation operators, and (ii) by definition, $\hat\rho_G$ and $\hat\rho$ share the same expectation value of quadratic operators.

The quantity (\ref{eq:quantity_not_good_for_calculation}) cannot be easily evaluated with tensor network techniques because it is generally hard to compute the von Neuman entropy $S(\hat\rho)$. Instead, the Rényi entropies $S_n(\hat\rho)=\frac{1}{1-n}\log{\rm tr}(\hat\rho^n)$ ---that reduce to the Von Neumann entropy in the replica limit $n\to1$--- are easy to represent as tensor network 
contractions for integer $n$. Therefore, in what follows we employ the Rényi-2 entropy $S_2=-\log{\rm tr}(\hat\rho^2)$ as a proxy. We use the R\'enyi-2 relative entropy as a quantifier of non-Gaussianity,
\be
    \label{eq:delta_rho_rho}
\delta(\hat\rho, \hat\rho_G) \equiv S_2(\hat\rho_G) - S_2(\hat\rho) = \log\frac{Z_2}{Z_2^G}\,,
\ee 
where $Z_2\equiv\rm tr(\hat{\rho}^2)$\, is the purity of $\hat{\rho}$. We also compare $\delta(\hat\rho, \hat\rho_G)$ with $\delta(\hat\rho,\hat\rho_{{\rm GGE}})$ to see how close the analytic solution $\hat\rho_{{\rm GGE}}$ is to the optimal Gaussian approximation~\footnote{Notice that the replacement ${\rm tr}(\hat\rho\log\hat\rho_{GGE}) \to {\rm tr}(\hat\rho_{GGE}\log\hat\rho_{GGE})$ is not exact for $S(\hat\rho||\hat\rho_{GGE})$, because $\hat\rho$ and $\hat\rho_{GGE}$ have different two-point functions. Nevertheless, the error that is made in the replacement is under control and we can still make it within our approximations: as we show in the Appendix \ref{A: non-gaussianity}, the error is of same the order of the distance between $\hat\rho_G$ and $\hat\rho_{GGE}$, which is a small correction to the distance between $\hat\rho$ and $\hat\rho_{GGE}$ if the largest component of the error is brought by non-Gaussianity, as we expect.}. 
In practice $Z_2$ is calculated from its tensor network representation, whereas $Z_2^G$ and $Z_2^{\rm GGE}$ are purities of two Gaussian states so they can be computed directly in terms of their correlation matrix  \cite{peschel2009reduced,surace2022}.

%

\subsection{Results for non-Gaussianity}

Our results for the non-Gaussianity of the exact state $\hat{\rho}(t)$ are reported in Fig.~\ref{fig: purity},  where we plot the relative entropy densities. The total non-Gaussianity developed in the process is small compared to the maximal value $L\log{2}$ \cite{lumia2024}, indicating that the $t$-GGE approximation provides a faithful description of the dynamics even beyond two-point functions. The accumulated error does not keep growing, but vanishes at long times as the steady state is captured exactly. Consistently, the distance of $\hat\rho$ from the reconstructed Gaussian partner $\hat{\rho}_G$ is smaller than its distance from the state $\hat{\rho}_{\rm GGE}$ obtained from the $t$-GGE equation. Nonetheless, the similarity of the two relative entropies indicates that $\hat\rho_{\rm GGE}$ is remarkably close to the Gaussian partner, and the trend with an initial accumulation of the error up to $\gamma t \sim 1$ is a feature of the optimal Gaussian approximation itself. 
%

Reducing the dissipation rate $\gamma$ improves the description in two simultaneous ways. First, as already observed, the rapidity distribution $\rho(t,k)$ provides a better approximation of the exact two-point functions, and this is now visible as the collapse of $\delta(\hat\rho,\hat\rho_{{\rm GGE}})$ onto $\delta(\hat\rho,\hat\rho_G)$. Second, the total non-Gaussianity is reduced together with error in higher correlation functions.
Notice that at a finite stationary density, the distance between the optimal Gaussian state and our Gaussian approximation is smaller than for $n_s=0$, besides the initial peak. This mirrors the behaviour of the rapidity distributions in Figs.~\ref{fig:summary_nk} and \ref{fig:nk_T}, which were almost exact for $n_s>0$. This is remarkable, since the larger total non-Gaussianity for $n_s>0$ does not spoil the accuracy of our approximation, as far as two-point functions are concerned. More on this behaviour is discussed in the Appendix \ref{A:Precision}.

\begin{figure*}
    \centering
    \includegraphics[width=\textwidth]{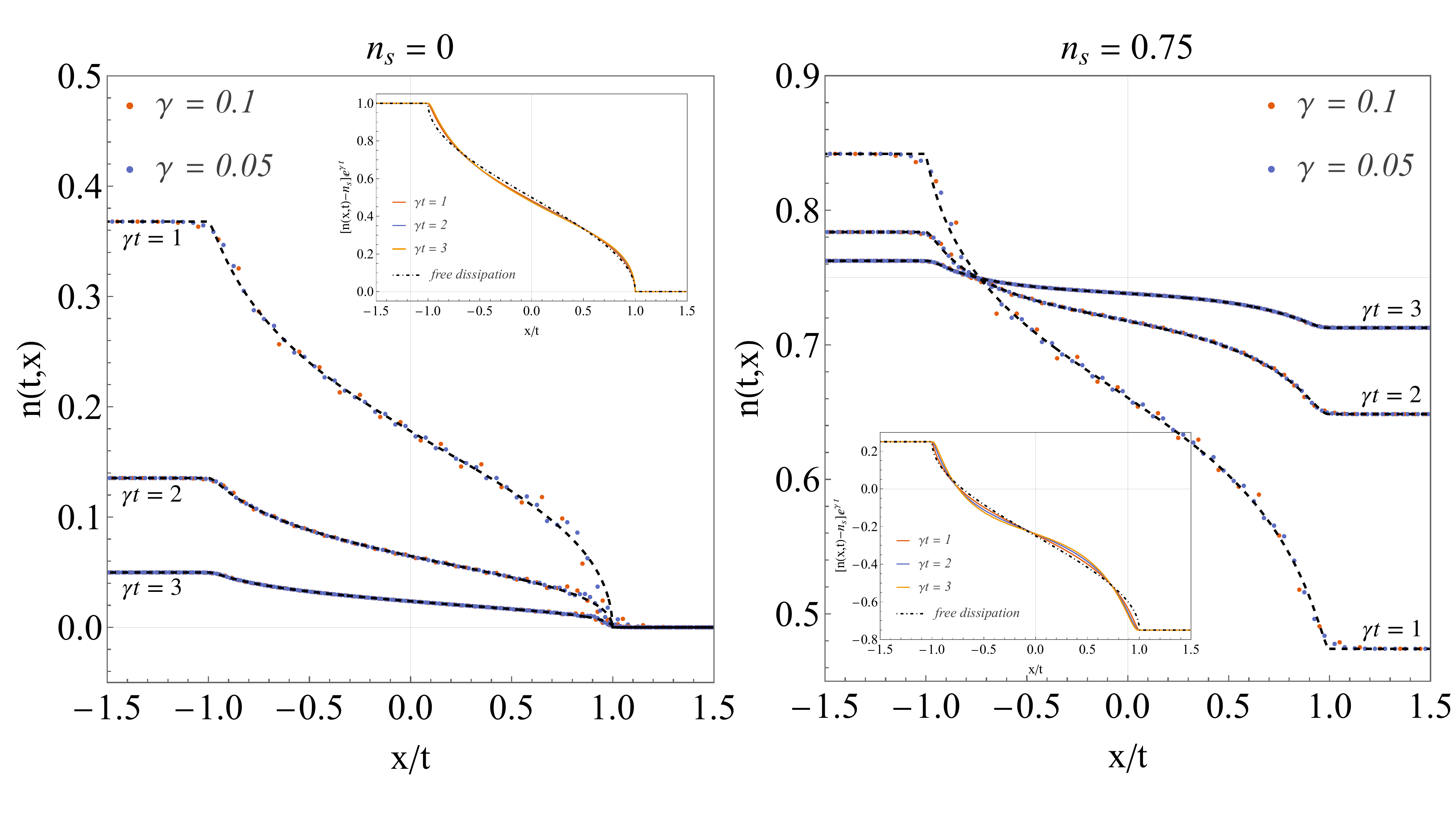}
    \caption{Comparison of the evolution of the density profile in real space $n(t,x)$ (dashed black lines) described by Eq.~\eqref{eq:evolution_transport} with initial density $n(0,x)=\theta(L/2-x)$ and simulation results, for $n_s=0$ (left) and $n_s=0.75$ (right). Dots represent the exact tensor network data for two different dissipation rates and system size values. We rescale the x-axis as $x/t$, so that the spreading of the excitation within the lightcone is fixed between $-1\le x/t\le1$. The finite-size dynamics collapses on the theoretical density in the appropriate limit. In the insert, a comparison between the profile for non-interacting losses (dot-dashed black line) and the solution of Eq.~\eqref{eq:evolution_transport} at different values of $\gamma t$ (full color lines). The $y$-axis is rescaled in a way that the $\gamma t$ dependence in the non-interacting (free) solution is removed, while bosonic losses carry a residual $\gamma t$ dependence. We refer to the main text for details.}
    \label{fig:transport}
\end{figure*}

\section{$\mathbf{t}-$GGE and transport}
\label{sec:inhomogeneous}

Now that we have validated the $t$-GGE approach for homogeneous conditions, let us turn to a situation with an inhomogeneous initial state, so that in addition to dissipation the system undergoes some kind of transport. We consider the evolution of an initial domain wall state, while particles are lost and added to the system. 
Within the dissipative GHD approach of Section~\ref{sec:tGGE-GHD}, the inhomogeneous dynamics is ruled by Eq.~\eqref{eq:evolution_transport}.

\subsection{Numerical solution of the dissipative GHD equation}

To solve Eq.~\eqref{eq:evolution_transport} numerically, we follow the split-step method described in Ref. \cite{riggio2023effects}, dividing the evolution between the pure transport part (l.h.s. of Eq.~\eqref{eq:evolution_transport}) and the dissipative part (r.h.s.) on each time step $t\to t+dt$. 
First, the phase space $(x,k)$ is discretized on a grid. We evolve the chosen initial profile $\rho(t,x,k)\to\rho'(t+dt,x,k)$ without the influence of dissipation for a time $dt$. The transport part is solved via the method of characteristics, hence the ancillary density $\rho'(t+dt,x,k)$ reads
\begin{equation}
\rho'(t+dt,x,k)=\rho(t,x-\sin(k)dt,k)\,.
\end{equation}
The result is defined on a new point $(x-\sin(k)dt,k)$, so the values on the original grid $(x,k)$ are obtained via interpolation. The GHD step can be interpreted as the underlying dynamics of quasi-particles, which evolve according to $\dot{x}=v(k),\,\dot{k}=-\partial_x V=0$ since there is no external potential. The resulting density $\rho'(t+dt,x,k)$ serves as initial condition for the master equation without transport, providing the evolved density $\rho'(t+dt,x,k)\to\rho(t+dt,x,k)$ as the solution of Eq.~\eqref{eq:self_consistent} at $t+dt$. The dissipative evolution between the times $t$ and $t+dt$ is obtained via a Runge-Kutta discretization of
\begin{equation}
    \partial_t \rho(t,x,k)=-\gamma \mathcal{F}[\rho(t,x,\cdot)](k)\,, 
\end{equation}
with $\rho(t+0,x,k)=\rho'(t+dt,x,k)$. The evaluation of the functional is efficient if we rely on the relation between the Hilbert transform and the Fourier transform (see Appendix \ref{A:Numerical ev of H}). By plugging back $\rho(t+dt,x,k)$ as initial condition for the next transport step, we can reconstruct the complete evolution of the rapidity distribution.

\subsection{Results}

We have performed MPO simulations of the dynamics of the hardcore boson gas with gain and losses for an initial domain-wall state. The evolution of the real-space density profile is displayed in Fig.~\ref{fig:transport}, for $n_s=0$ and $n_s=0.75$. In our simulations we use $(L=100,\gamma=0.1)$ and $(L=200$, $\gamma=0.05)$. We study the dynamics until $t=L/2$, right before the front bounces off the boundaries. 

For the numerical integration of  Eq.~\eqref{eq:evolution_transport}, we use the split-step procedure sketched above. As already mentioned in Sec.~\ref{sec:tGGE-GHD}, for an initial domain-wall state, the solution depends only on $\tau=\gamma t$ and $x/t$, so the size of the system does not matter provided that we properly rescale all quantities. In practice we use a total length $1$, $\gamma=10$, and a time step $dt=10^{-3}$, with a grid of $1000$ points in real-space and $200$ points in rapidity space for each value of $n_s=0,\,0.75$.
  
In Fig.~\ref{fig:transport}, we compare the density $n(t,x) = \int dk/2\pi \rho(t,x,k)$ obtained from the numerical solution of Eq.~\eqref{eq:evolution_transport} with our tensor network simulations. We see that the integrated equation provides an extremely accurate description of the dynamics, indicating that the GHD picture remains valid even when subject to dissipation. The errors visible for short times are not induced by dissipation but are a consequence of pure GHD, which holds exactly only in Euler's large space-time scales. As discussed in the section \ref{sec:tGGE-GHD}, the accuracy of our description is further confirmed by the collapse onto the theoretical solution, that we observe for increasing $L$ and decreasing $\gamma$ at a fixed $\gamma L$, where the Euler limit and the slow dissipation limit are simultaneously approached.

Notice that the density profile predicted by dissipative GHD is not trivial, in the sense that it is different from a simple decay of the pure GHD without dissipation. To elaborate,  without dissipation the evolution of the profile predicted by GHD would be $n_{\gamma = 0}(t,x)=1/2-\arcsin(x/t)/\pi$ within the lightcone $-1\le x/t \le 1$ \cite{antal1999profile}. Then, if the dissipative evolution consisted of adding and removing {\it non-interacting} particles (as in the model of Refs.~\cite{alba2021spreading,alba2023logarithmic}, with gain and loss of non-interacting fermions), then the population of each mode with rapidity $k$ would evolve independently of all other modes and would relax exponentially to the stationary density $n_{s}$. Then the particle density would simply be given by $n(t,x)=n_s + (n_{\gamma =0}(t,x)-n_s)\,e^{-\gamma t}$. However, in Fig.~\ref{fig:transport} we show that this is {\it not} what is happening here, and the density profile does differ significantly from the non-interacting case. As highlighted in Section~\ref{sec:tGGEreview}, the difference arises from the bosonic nature of the gain and losses and from the associated Jordan-Wigner strings that make the loss and gain functionals $\mathcal{F}_{L,G} [\rho]$ non-local in rapidity space. Consistently, Fig.~\ref{fig:transport} shows that this difference becomes more pronounced for larger $\gamma t$, as the number of gain and loss processes that have taken place is increased.

\section{Conclusion}\label{sec:conclusion}

In this work, we investigated the open dynamics of integrable quantum systems under weak dissipation. For concreteness, we focused on a model of lattice hardcore bosons subject to slow atom loss and gain processes.
Even if the dissipative processes break integrability, the evolving state of the system can be described employing the $t$-GGE approach, which assumes fast local relaxation to a local GGE and which yields a closed equation for the evolution of the rapidity distribution that parameterizes the GGE.
This method allowed us to understand how conserved quantities evolve in time, providing an accurate approximation for the non-equilibrium dynamics, that becomes exact in the limit of adiabatic dissipation.
Despite being a simplified description, the $t$-GGE accurately tracks the evolution of the system, also capturing the convergence towards the exact stationary state, which, for hardcore bosons, is a Gaussian state of the underlying Jordan-Wigner fermions.

Our study highlights the effectiveness of the $t$-GGE approach in reproducing both single-particle observables, e.g. local densities in real and momentum space, as well as higher correlations, which have been compactly inspected via a non-Gaussianity measure. By comparing the exact state $\hat\rho$, obtained using tensor network simulations, with the $t$-GGE approximation $\hat\rho_{{\rm GGE}}$, we observed that, although slight deviations initially accumulate, the error stabilizes and diminishes as the system loses coherence and moves towards equilibrium. In this way, we are able to capture analytically the optimal Gaussian approximation of the dynamics without having to resort to expensive variational approaches.

The dissipation functional obtained acting on the $t$-GGE can be used in combination with Generalized Hydrodynamics to describe transport properties in an open system. The combined description implies peculiar scaling properties of the exact solution, that we show to be correct besides for finite-size effects. In this way, we show that the time-dependent GGE provides an accurate description of the dynamics for both homogeneous and inhomogenous conditions.

\acknowledgments{We thank  Leonardo Mazza, Dragi Karevski, Zala Lenarčič, Vincenzo Alba and Iris Ulčakar for useful discussions. This work was supported by the PNRR MUR project PE0000023-NQSTI, the PRIN 2022 (2022R35ZBF) - PE2 - ``ManyQLowD'', and by the Agence Nationale de la Recherche through ANR-20-CE30-0017-02 project
‘QUADY’ and ANR-22-CE30-0004-01 project ‘UNIOPEN’.}

\appendix
 \section{The gain functional \label{A:GainFun}}

In this appendix, we complete the derivation of the gain functional in the thermodynamic limit as sketched in Section~\ref{sec:homogeneous}. We first recap a few well-known facts about hardcore bosons with the purpose of fixing the notation before discussing the details of the gain functional.

\subsection{Diagonalization of hardcore bosons}
The Hamiltonian of hopping hardcore bosons
\be
\hat H = -\frac{1}{2} \sum_{j=1}^L \left( \hat\sigma_{j}^{+}  \hat\sigma_{j+1}^{-}
+\hat\sigma_{j}^{-}  \hat\sigma_{j+1}^{+} \right)\,,
\ee
under the Jordan-Wigner mapping $\hat c_j = \prod_{i<j} \hat \sigma^{z}_{i} \; \hat\sigma^{+}_{j}$ is transformed into a free fermionic model
\be
\hat H_{JW} = -\frac{1}{2} \sum_{j=1}^L \left( \hat{c}_{j}^{\dagger}  \hat{c}_{j+1}
+\hat{c}_{j+1}^{\dagger}  \hat{c}_{j} \right)\,.
\ee
The boundary terms are mapped to $\hat\sigma_{L}^+\hat\sigma_{1}^-+\hat\sigma_{1}^+\hat\sigma_{L}^- \,\to\,-e^{i\pi\hat{N}} (\hat{c}_{L}^{\dagger} \hat{c}_{1}
+\hat{c}_{1}^{\dagger}  \hat{c}_{L})$, so that PBC on the original Hamiltonian become antiperiodic or periodic on fermions depending on the total number of particles:
\begin{equation}
\label{eq: bc}
\hat{c}^\dagger_{L+1}=-e^{i\pi\hat{N}}\hat{c}^\dagger_1\,.
\end{equation}
Free fermions are diagonalized by Fourier modes
\begin{equation}
    \hat c^\dagger(k)=\dfrac{1}{\sqrt{L}}\sum_{j=1}^Le^{ikj}\hat c_j^\dagger\,,
\end{equation}
where the set of allowed rapidities is fixed by the boundary conditions in Eq.~\eqref{eq: bc}. If $N$ is even $k\in Q_{ap}$, while for $N$ odd $k \in Q_p,$ where
\begin{gather}
    Q_p=\dfrac{2\pi}{L}\{1,\dots,L\}\,,\\
    Q_{ap}=\dfrac{2\pi}{L}\{1/2,\dots,L-1/2\}\,.
\end{gather}
Now $\hat n(k)=\hat c^\dagger(k) \hat c(k)$ diagonalize $\hat H = \sum_{k} \epsilon(k) \hat n(k)$, $\epsilon(k)=-\cos(k)$, so that $\hat c^\dagger(k)$ creates conserved quasi-particles and $\{\hat{n}(k)\}_{k\in Q_{p/ap}}$ is the set of integrals of motion in involution of this simple integrable model.

\subsection{Derivation of the functional}
First, we notice that the Lindbladian $\mathscr{L}$ satisfies
\begin{equation}
    [\hat{N},\mathscr{L}\hat\rho\,] = \mathscr{L}([\hat{N},\rho\,])\,,
\end{equation}
i.e. the superoperators $\mathscr{L}$ and $[\hat{N},{\cdot}\,]$ commute. This property is called a \emph{weak symmetry}: $\hat{N}$ is not conserved in time but $\mathscr{L}$ is still block diagonal and
\begin{equation}
    \frac{d}{dt} [\hat{N},\hat{\rho}\,]=\mathscr{L}([\hat{N},\hat\rho\,])\,,
\end{equation}
therefore, a state with an initially defined number of particles will commute with $\hat{N}$ at any time. As a consequence, one can always decompose $\hat{\rho}=\bigoplus_N \hat{\rho}^{(N)}$ where $\hat{\rho}^{(N)}$ has support only on states with a fixed number of particles. We only need to split it into the parity sectors \begin{equation}
    \hat{\rho} = \hat{\rho}^{(e)} \oplus \hat{\rho}^{(o)}\,.
\end{equation}
The Hamiltonian dynamics conserves the parity, but not the full dissipative dynamics. A state-supported on both parity sectors has finite occupations on both sets of rapidities $Q_p,\,Q_{ap}$ and the conserved charges are
\begin{equation*}
    \hat{Q}(k) = \hat{P}_+ \delta_{k\in Q_{ap}}\hat{c}^\dagger(k)\hat{c}(k) \hat{P}_+ + \hat{P}_- \delta_{k\in Q_{p}}\hat{c}^\dagger(k)\hat{c}(k) \hat{P}_- 
\end{equation*}
so that the functional \eqref{eq:F_gauss G} should be regarded as
\begin{equation}
    \mathcal{F}_{G}[\rho_g](k)=\sum_{j=1}^L \text{Re\,}\langle \Hat L_{j}[\Hat L_{j}^\dagger,\Hat Q(k)]\rangle_{\hat{\rho}^{(e)} \oplus \hat\rho^{(o)}}\,,
\end{equation}
where we introduced $\hat{P}_\pm=\frac{1}{2}(1\pm(-1)^{\hat{N}})$. Using the translational invariance of the system, one finds
\begin{equation}
    \mathcal{F}_G
    [\rho](k) =L\langle \Hat\sigma_1^+\Hat\sigma_1^-\Hat Q(k)\rangle_{\hat\rho}
    -L\langle \Hat\sigma_1^+\Hat Q(k)\Hat\sigma_1^-\rangle_{\hat\rho}.
    \label{eq:F rhoG 1 and 2}
\end{equation}
Consider now, for instance, $k\in Q_{ap}$. Then
\begin{align}
    \langle \hat{\sigma}_1^+\hat{\sigma}_1^- \hat Q(k)\rangle_{\hat\rho} &= \text{tr\,}[\hat{\rho} \hat{\sigma}_1^+\hat{\sigma}_1^- \hat{P}_+\hat{c}^\dagger(k)\hat{c}(k)\hat{P}_+] \\
    &=\text{tr\,}[\hat{P}_+\hat{\rho}\hat{P}_+ \hat{\sigma}_1^+\hat{\sigma}_1^-\hat{c}^\dagger(k)\hat{c}(k)]
\end{align}
for the cyclic property of the trace. We can reduce the calculation on the single sector $\hat{\rho}^{(e)}=\hat{P}_+\hat{\rho}\hat{P}_+$. Introducing $\hat \sigma_1^-=\hat c_1^\dagger=\frac{1}{\sqrt{L}} \sum_q e^{iq}\hat c^\dagger(q)$, now 
%
%
%
%
%
%
%
%
\begin{align*}
    L \langle \sigma_1^+ & \hat \sigma_1^-\hat c^\dagger(k)\hat c(k)\rangle_{\hat\rho^{(e)}} =L\langle \hat c_1\hat c_1^\dagger \hat c^\dagger(k)\hat c(k)\rangle_{\hat\rho^{(e)}}\nonumber\\
    &=\sum_{qq'} e^{i(q-q')} \langle \hat c(q')\hat c^\dagger(q) \hat c^\dagger(k)\hat c(k)\rangle_{\hat\rho^{(e)}}
    \nonumber\\
    &=\sum_{qq'} e^{i(q-q')} \Bigl( \langle \hat c(q')\hat c^\dagger(q) \rangle \langle \hat c^\dagger(k)\hat c(k)\rangle \nonumber \\
    & \qquad \qquad \qquad - \langle \hat c(q')\hat c^\dagger(k)\rangle\langle \hat c^\dagger(q)\hat c(k)\rangle \Bigr)
    \nonumber\\
    &=L \rho(k)-\langle \hat N\rangle  \rho(k) - \rho(k)[1- \rho(k)],
\end{align*}
where we have used Wick's theorem for Gaussian states. For the second term, more care is required because $\hat{c}^\dagger(k)\hat{c}(k)$ is inserted between $\hat \sigma_1^+$ and $\hat \sigma_1^-$, which changes the parity of particles. Employing $\hat P_+\hat \sigma_1^\pm=\hat \sigma_1^\pm \hat P_-$,
\begin{align*}
    \langle \hat{\sigma}_1^+ Q(k)\hat{\sigma}_1^-\rangle_{\hat\rho} &= \text{tr\,}[\hat{\rho} \hat{\sigma}_1^+ \hat{P}_+\hat{c}^\dagger(k)\hat{c}(k)\hat{P}_+\hat{\sigma}_1^-] \\
    &=\text{tr\,}[\hat{P}_-\hat{\rho}\hat{P}_- \hat{\sigma}_1^+\hat{\sigma}_1^-\hat{c}^\dagger(k)\hat{c}(k)]\,.
\end{align*}
If $k\in Q_{ap}$, the average is computed over $\hat{P}_-\hat{\rho}\hat{P}_-=\hat{\rho}^{(o)}$ of the opposite sector and we rely on the identity 
%
%
%
%
\begin{equation}
    \hat c(k)=\dfrac{i}{L}\sum_{q\in Q_p}\dfrac{e^{i(q-k)/2}}{\sin((q-k)/2)}\,\hat c(q)\,.
\end{equation}
An analogous formula holds in the complementary case. Again, we develop the term using Wick's theorem
\begin{align*}
    L\langle \hat \sigma_1^+\hat c^\dagger(k)\hat c(k)\hat \sigma_1^-\rangle&=\dfrac{1}{L}\sum_{q,q'}\dfrac{e^{i(q-q')/2}}{\sin\frac{q-k}{2}\sin\frac{q'-k}{2}}\langle \hat c_1 \hat c^\dagger(q')\hat c(q)\hat c_1^\dagger\rangle\nonumber\\
    &=\underbrace{\dfrac{1}{L}\sum_{q,q'}\dfrac{e^{i(q-q')/2}}{\sin\frac{q-k}{2}\sin\frac{q'-k}{2}}\langle \hat c_1 \hat c^\dagger(q')\rangle\langle \hat c(q)\hat c_1^\dagger\rangle}_{(\star)}\nonumber\\
    &+\underbrace{\dfrac{1}{L}\sum_{q,q'}\dfrac{e^{i(q-q')/2}}{\sin\frac{q-k}{2}\sin\frac{q'-k}{2}}\langle \hat c_1\hat c_1^\dagger \rangle\langle \hat c^\dagger(q')\hat c(q)\rangle}_{(\star \star)}.
\end{align*}
In the following, we treat the two terms separately, starting from the simpler $(\star\star)$ 
\begin{align}
    (\star\star)&=\dfrac{1}{L}\sum_{q,q'}\dfrac{e^{i(q-q')/2}}{\sin\frac{q-k}{2}\sin\frac{q'-k}{2}}\langle [1-\hat c_1^\dagger \hat c_1] \rangle\langle \hat c^\dagger(q')\hat c(q)\rangle\nonumber\\
    &=\dfrac{1}{L}\sum_{q}\dfrac{\rho(q)}{\sin^2\frac{q-k}{2}}\langle [1-\hat c_1^\dagger \hat c_1] \rangle\nonumber\\
    &=\dfrac{1-\langle \hat N \rangle/L}{L}\sum_{q}\dfrac{ \rho(q)}{\sin^2\frac{q-k}{2}}\,.
\end{align}
%
Analogously, we have for $(\star)$
\begin{align*}
    (\star)&=\dfrac{1}{L^2}\sum_{q,q'}\dfrac{e^{i(q-q')/2}}{\sin\frac{q-k}{2}\sin\frac{q'-k}{2}}\sum_{p,p'}e^{i(p-p')}
    \nonumber\\&\times\langle \hat c(p) \hat c^\dagger(q')\rangle\langle \hat c(q)\hat c^\dagger(p')\rangle\nonumber\\
    &=\dfrac{1}{L^2}\sum_{q,q'}\dfrac{e^{-i(q-q')/2}}{\sin\frac{q-k}{2}\sin\frac{q'-k}{2}}\left( 1- \rho(q')\right)\left( 1-  \rho(q)\right)\nonumber\\
    &=\dfrac{1}{L^2}\sum_{q,q'}\left[\cot{\dfrac{q-k}{2}}\cot{\dfrac{q'-k}{2}}+1\right]
    \nonumber\\&\times\left( 1-  \rho(q')\right)\left( 1-  \rho(q)\right)\nonumber\\
    &=\left(1-\dfrac{\langle \hat N\rangle}{L}\right)^2+\left[\dfrac{1}{L}\sum_q\cot{\left(\dfrac{q-k}{2}\right)}(1- \rho(q))\right]^2.
\end{align*}
Gathering the two terms $(\star)$, and $(\star\star)$, we get
\begin{align}
    \mathcal{F}_G[\rho](k)=-&\left[1- \rho(k)\right] \rho(k) \nonumber \\
    -&\dfrac{1-\langle\hat N\rangle/L}{L}\sum_q\dfrac{ \rho(q)- \rho(k)}{\sin^2\frac{q-k}{2}} \label{eq:GainFun any L} \\
    -\left(1-\dfrac{\langle \hat N\rangle}{L}\right)^2-&\left[\dfrac{1}{L}\sum_q\cot{\left(\dfrac{q-k}{2}\right)}(1- \rho(q))\right]^2\nonumber\,,
\end{align}
\noindent where we have used the identity $\sum_q1/\sin^2(\frac{q-k}{2})=L^2$ to reduce the degree of the pole in the second term in Eq.\eqref{eq:GainFun any L}. Moving to the thermodynamic limit, we obtain the final expression

\begin{align}
    \mathcal{F}_G&[\rho_g](k)=-\rho(t,k)+\rho^2(t,k)-\mathcal{H}^2\left[\rho(t,\cdot)\right](k)\nonumber\\
    &-\left(1- n(t)\right)^2+2(1- n(t))\mathcal{H}'\left[\rho(t,\cdot)\right](k)\,, 
\end{align}
where we have used $\mathcal{H}[const]=0$ and the derivative of the Hilbert transform is define as $\mathcal{H}'\left[f\right](k)=\frac{1}{\pi}P.V.\int dz\frac{f(k)-f(z)}{(k-z)^2}$. The expression of the derivative has been subtracted from its Hadamard finite part to regularize the nonlinearity in $k=z$.

\subsection{Solving the master equation\label{A:MastEq SignalF}}
 This appendix will shortly present a method to solve the evolution equation of the rapidity distribution Eq. \eqref{eq:effective eq}. Following the same procedure described in \cite{riggio2023effects}, we can simplify the master equation by shifting the problem to $Q(t,z)$, the signal function associated with $\rho(t,k)$. This analytic function is constructed by adjoining, to $\rho(t,k)$, an imaginary part under the form of its Hilbert transform. To benefit from this perspective shift, one has to consider the analytical continuation of $Q(z)$ to the complex plane \cite{GrafakosClassicalFourierAnalysis}: $Q(\overline{z})=i\mathcal{H}[\rho]$, which in case of $2\pi$ periodic $\rho$ is valid for $\text{Im}(z)>0$ and $\text{Re}(z)\in[-\pi,\pi]$.


We observe that in the complex plane, the real part of $Q(z)$ is absorbed into the Hilbert transform (so into its imaginary part). Hence, the real and imaginary parts are no longer separated. The procedure is then simple, we have to consider $\eqref{eq:effective eq}+i\mathcal{H}[\eqref{eq:effective eq}]$ to build an effective master equation for $Q(\overline{z})$ which real part of the solution boils down to $\rho(t,k)$. For conciseness, we will eliminate the $t$, and $k$ dependences on $\rho$ and $Q$ and drop the overline of $z$ for the rest of this appendix.

The master equation for single particle losses has already been obtained in \cite{riggio2023effects}, so we now search for its gain equivalent. First, let us recall the evolution equation of $\rho$ for single particles gains
\begin{align}
\partial_t{\rho}=-\gamma_G&\big\{-\rho+\left(\rho^2-\mathcal{H}^2\left[\rho\right]\right)\nonumber\\&-\left(1-n(t)\right)^2+2(1-n(t))\mathcal{H}'\left[\rho\right]\big\}\,.
\label{eq:MastEq rho_g Gain}
\end{align}
\noindent As explained, the next step is to consider \eqref{eq:MastEq rho_g Gain}+$i\mathcal{H}$[\eqref{eq:MastEq rho_g Gain}], which gives 
\begin{align}
    &\underbrace{\partial_t{\rho}+i \partial_t{\mathcal{H}[\rho]}}_{=\overset{.}{Q}}=-\gamma_G\bigg\{\underbrace{-\rho-i\mathcal{H}[\rho]}_{=Q}\nonumber\\&
    +\underbrace{\left(\rho^2-\mathcal{H}^2[\rho] \right)+i\mathcal{H}\left[\rho^2-\mathcal{H}^2[\rho_g]\right]}_{=X}-
    (1-n(t))^2\nonumber\\&
    +\underbrace{2(1-n(t))\mathcal{H}'[\rho_g]+2i(1-n(t))\mathcal{H}\left[\mathcal{H}'(\rho)\right]}_{=Y}\bigg\}.
\end{align}
The term $Y$ is easily obtained thanks to Hilbert transform properties: $\mathcal{H}[\mathcal{H}(f)]=-f$, and $\mathcal{H}'[f]=\mathcal{H}[f']$, so that
\begin{align}  
Y&=-2i(1-n(t))\partial_zQ,
\end{align}
\noindent For $X$, we have to analyze the square of a signal function briefly. Since, $Q^2(z)$ is analytic for $\text{Im(z)}>0$, the function $\rho^2-\mathcal{H}[\rho]^2+2i\rho\mathcal{H}[\rho]$ is an analytic signal if and only if $\mathcal{H}[\rho^2-\mathcal{H}[\rho]^2]=2\rho\mathcal{H}[\rho]$. It follows that $X=Q^2$. Finally, gathering all the terms, we obtain an equation of evolution for $Q$
\begin{equation}
    \partial_t Q=-\gamma_G\left[-Q+Q^2-2i(1-n(t))\partial_z Q-(1-n(t))^2\right].
\end{equation}
We can now collect the term with respective losses and gains to recover the equivalent of Eq.\eqref{eq:effective eq} in terms of an analytic function, precisely
\begin{widetext}
\begin{align}
    \partial_t{Q}=&-\gamma_L\left[Q-Q^2-2in(t)\partial_z Q+n^2(t)\right]\nonumber\\
    &-\gamma_G\left[-Q+Q^2-2i(1-n(t))\partial_z Q-(1-n(t))^2\right]
\end{align}
\end{widetext}
%

%
\noindent Consistently, this equation admits a solution for any $t$ and $\gamma=\gamma_L+\gamma_G$ positive. This equation can be further reduced by moving from $\gamma_{G,L}$ to $\gamma$ and $n_s$. This leads to the ensuing condensed expression

\begin{align}
    \partial_\tau Q(\tau,z')=&(2n_s-1)\left(Q(\tau,z')-Q^2(\tau,z')
    +n^2(\tau)\right)\nonumber\\&-2n_s(n(\tau)-1/2) \label{eq:reduced Masteq}
\end{align}

\noindent with $\tau=\gamma t$, and $z'=z+g(t)$, with $g(t)=2i\left(e^{-\tau}-1\right)(n_0-n_s)(2n_s-1)+4in_s(1-n_s)\tau$. We can now find a general equation for Eq. \eqref{eq:reduced Masteq}. For this, we exploit the structure of the mean density to guess the final form of the solution. Indeed, one can suppose $Q(\tau,z')=n_s+\alpha(\tau,z')e^{-\tau}$ and replace the expression of $Q(\tau,z')$ in Eq. \eqref{eq:reduced Masteq}. Hence,

\begin{align}
     \dfrac{\partial_\tau \alpha}{(n_0-n_s)^2-\alpha^2}+\dfrac{4n_s(n_s-1)}{(n_0-n_s)+\alpha}=(2n_s-1)e^{-\tau}.
\end{align}

\noindent This equation can be exactly solved, and a general solution could be expressed as

\begin{equation}
    \alpha(\tau,z')=(n_0-n_s)+\dfrac{e^{2e^{-\tau}(n_0-n_s)(2n_s-1)+\tau}}{\Tilde{E}(\tau)+c(z')e^{(2n_s-1)^2\tau}},
\end{equation}

\noindent where the function $\Tilde{E}(\tau)$ reads

\begin{equation}
    \Tilde{E}(\tau)=(2n_s-1)E_{4n_s(n_s-1)}\left(2e^{-\tau}(n_s-n_0)(2n_s-1)\right),
\end{equation}

\noindent and is defined via the generalized exponential function $E_n(x)=\int_1^\infty~dt e^{-xt}/t^n$. We can easily extract $c(z')$ by recalling that at $\tau=0$, $Q(0,z')=Q_0(z')=n_s+\alpha(0,z')$ which leads to

\begin{multline}
    \alpha(\tau,z')=(n_0-n_s)\\+\dfrac{e^{2e^{-\tau}(n_0-n_s)(2n_s-1)+\tau}}{\Tilde{E}(\tau)+\left(\dfrac{e^{2(n_0-n_s)(2n_s-1)}}{n_0-Q_0(z')}+\Tilde{E}(0)\right)e^{(2n_s-1)^2\tau}}.
\end{multline}

\noindent To explicit the function $Q_0(z')$ we can rely on the definition $Q(z')=\frac{i}{2\pi}\int_{^-\pi}^\pi\frac{dq~\rho(q)}{\tan\left(\frac{z'-q}{2}\right)}$ of the analytic signal. It follows

\begin{multline}
    \alpha(\tau,z')=(n_0-n_s)\\+\dfrac{e^{2e^{-\tau}(n_0-n_s)(2n_s-1)+\tau}}{\Tilde{E}(\tau)+\left(\dfrac{e^{2(n_0-n_s)(2n_s-1)}}{n_0-i\mathcal{H}[\rho](z')}+\Tilde{E}(0)\right)e^{(2n_s-1)^2\tau}},
\end{multline}

\noindent where $\rho(z)$ is the chosen initial density profile. By rewriting the generalized exponential function in terms of $g(t)$ we end up with the Eq. \eqref{eq:analytic_solution}

\subsection{Numerical evaluation of the functional \label{A:Numerical ev of H}}
The dissipation functional must be evaluated numerically to solve the effective equation with transport Eq.~\eqref{eq:evolution_transport}, as we have discussed in Sec.~\ref{sec:inhomogeneous}. Below, we explain an efficient approach to perform the calculation.
The previous expression Eq.~\eqref{eq:GainFun any L} provides the natural lattice discretization of the Hilbert transform
\begin{equation}
    \mathcal{H}f(k_j) = \dfrac{1}{L}\sum_{q_i}\cot{\left(\dfrac{q_i-k_j}{2}\right)} f(q_i)\,,
\end{equation}
where if $k_j\in Q_{p}=\{2\pi/L\cdot j\}$ then $q_i\in Q_{ap}=\{2\pi/L\cdot(j-1/2)\}$ and viceversa. Notice that the separation of lattices between $k$ and $q$ naturally avoids the singularity, so that the thermodynamic limit correctly provides the Cauchy principal value. Suppose that $k_j=2\pi/L\cdot j$. The lattice Hilbert transform can be related to a discrete Fourier transform by noticing that
\begin{align}
    DFT[\mathcal{H}f](m) &= \sum_{j=0}^{L-1} e^{-i k_j m} \mathcal{H}f(k_j) \\
    & = -i e^{i{\pi m}/{L}} DFT[f](m) 
\end{align}
if $m>0$, $=0$ if $m=0$. Inverting this expression provides an efficient approach for computing  $\mathcal{H}f$ provided that we use a fast Fourier transform.

\section{Non-Gaussianity}
\label{A: non-gaussianity}

In this appendix, we discuss the impact of the approximation
$\tr(\hat\rho\log\hat\rho_{GGE})\approx\tr(\hat\rho_{GGE}\log\hat\rho_{GGE})$ made in Section \ref{sec:homogeneous}, proving that is a minor correction if the largest component of the error is brought by non-Gaussianity. 

Under this hypothesis, the distance between $\hat\rho_G$ and $\hat\rho_{GGE}$ is smaller than the distance between $\hat\rho_{GGE}$ and $\hat\rho$. We set therefore $\hat\rho\equiv\hat\rho_{GGE}+\hat\epsilon$, $\hat\rho_{GGE}\equiv \hat\rho_G + \hat\delta$ with $\norm{\hat\delta}_2 \ll \norm{\hat\epsilon}_2$, where $\norm{\hat A}_2=\tr (\hat{A}^\dagger \hat{A}\,)$ is the Hilbert-Schmidt norm. 
For a matrix $\hat{A}$ having small norm of order $O(\norm{\delta}_2)$ we directly denote $\hat{A}=O(\delta)$. We will use the following properties:
\begin{enumerate}[label=(\roman*)]
    \item $\,\,\norm{\hat{A}\hat{B}}_2\le\norm{\hat{A}}_2\norm{\hat B}_2$
    \item $\,\,\log(1+\hat{A})=O(\delta)\,$ if $\,\hat A = O(\delta)$
    \item $\,\,\tr{\hat A}=O(\delta)\,$ if $\,\hat A = O(\delta)$.
\end{enumerate} 
(i) is a standard property of the Hilbert-Schmidt norm, while (ii) is a trivial consequence of the Taylor series expansion for the matrix logarithm. The last property can be proved by noting that $|\tr\hat A | \le \sum_i |\lambda_i|$ if $\lambda_i$ are the eigenvalues of $\hat A$. For any eigenvalue $|\lambda_i|=\norm{\hat A |v_i\rangle}$ if we normalize the associated eigenvector $|v_i\rangle$. Then $|\lambda_i| \le \rm{sup}_{| \lVert v\rangle \rVert =1} \norm{A |v\rangle}= \norm{\hat A}_{op}$, that is the operator norm of $\hat{A}$. Expanding a normalized $|v\rangle$ on an orthonormal basis leads to $\norm{\hat A|v\rangle} = \norm{\sum_j v_j \hat A |j\rangle}\le\sum_j|v_j|\,\norm{\hat A |j\rangle}$. By the Cauchy-Schwarz inequality $\norm{\hat{A}|v\rangle}\le\bigl(\sum_j |v_j|^2 \bigr)^{1/2}\bigl(\sum_i \norm{\hat A |j\rangle}^2 \bigr)^{1/2}=\norm{\hat A}_2$, resulting in $\norm{\hat A}_{op} \le \norm{\hat A}_2$. Overall we have $|\tr{\hat A}| \le D\norm{\hat A}_{op}\le D\norm{\hat A}_2$, that for a system with fixed Hilbert space dimension $D$ implies that the trace is $O(\delta)$. 
Getting back to the $\hat\rho_{GGE}$ replacement, let us expand 
\begin{equation*}
    {\rm tr}(\hat\rho\log\hat\rho_{GGE}) = {\rm tr}\Bigl\{\hat\rho\log\Bigl[\hat\rho_G(\mathbb{I}+\hat\rho_G^{-1}{\hat\delta})\Bigr]\Bigr\}\,.
\end{equation*}
The logarithm of a matrix product can be transformed into a sum only when the matrices commute, otherwise, the BCH formula implies $\log(\hat A \hat B)=\log\hat A + \log\hat B + \frac{1}{2}[\log\hat A, \log\hat B]+...$\,. Here $\hat{A}=\hat \rho_G,\,\hat{B}=\mathbb{I}+\hat\rho_G^{-1}{\hat\delta}$, so that $\log\hat B = O(\delta)$ and the commutator is $O(\delta)$. Then
\begin{equation*}
    {\rm tr}(\hat\rho\log\hat\rho_{GGE}) = \tr{(\hat\rho\log\hat\rho_G)}+\tr{\bigl[\hat\rho\log(\mathbb{I}+\hat\rho_G^{-1}\hat\delta]\bigr)} + O(\delta)\,.
\end{equation*}
The second term is also an $O(\delta)$ error by the properties above, while we can replace exactly $\hat\rho\to\hat\rho_G$ in the first term. Expanding again $\hat\rho_G = \hat\rho_{GGE}-\hat\delta$ we find
\begin{equation*}
\begin{split}
    {\rm tr}(\hat\rho\log\hat\rho_{GGE}) & = \tr(\hat\rho_{GGE}\log\hat\rho_{GGE}) \\
    & \quad + \tr\bigl[\hat\rho_{GGE} \log(\mathbb{I}-\hat\rho_{GGE}^{-1}\hat\delta)\bigr]\\
    &\quad -\tr\bigl[\hat\delta\log(\hat\rho_{GGE}-\delta)]+O(\delta)\,\\
    & =  \tr(\hat\rho_{GGE}\log\hat\rho_{GGE}) + O(\delta)\,, 
\end{split}
\end{equation*}
so the replacement $\hat\rho\to\hat\rho_{GGE}$ is indeed a small error compared to the $O(\epsilon)$ distance we are considering, if the error is mainly due to non-Gaussianity.

\begin{figure*}[t]
\centering
    \includegraphics[width=0.9\textwidth]{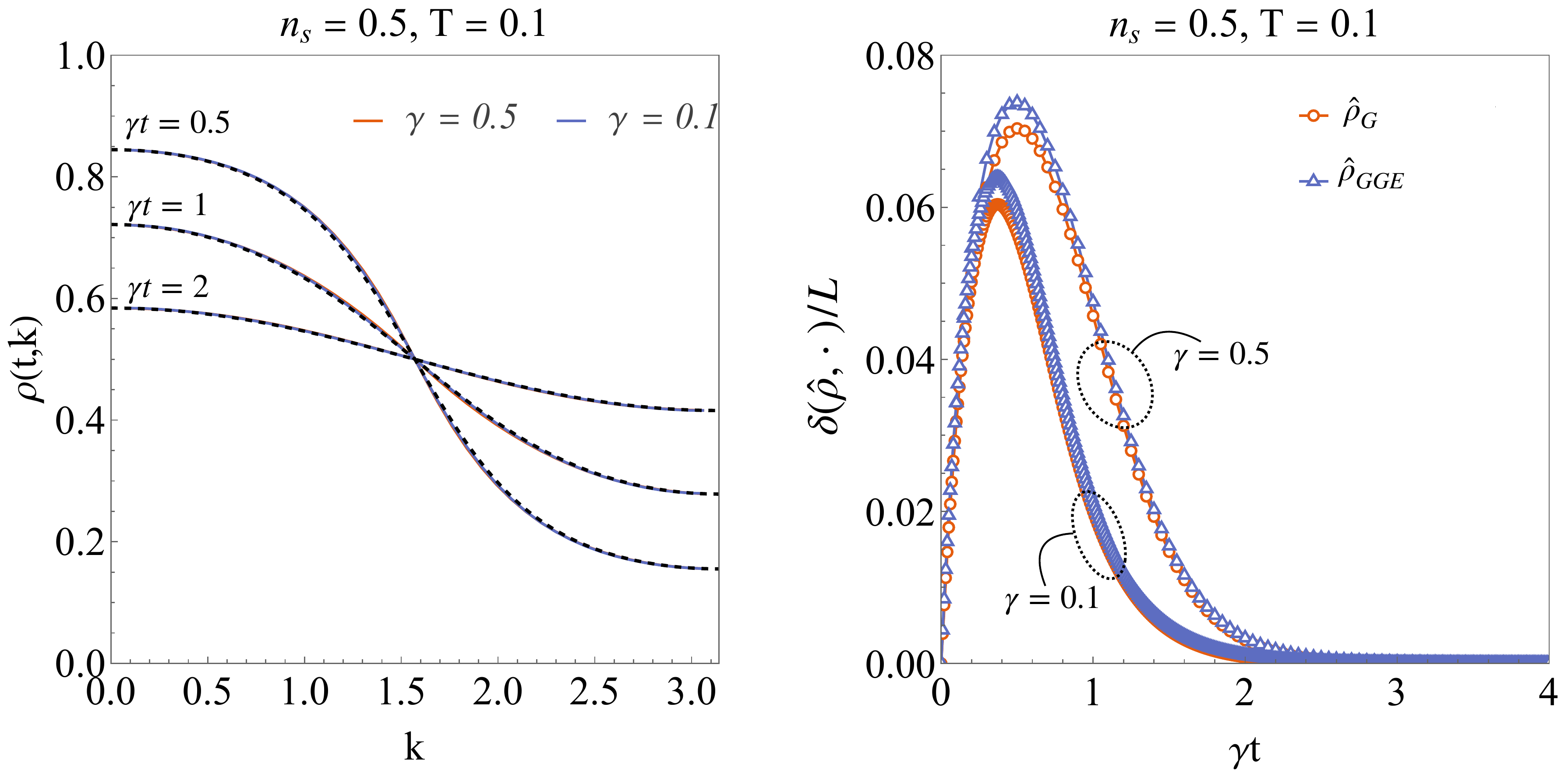}
\caption{Balanced losses and gains at $n_s = 1/2$.
(left) Comparison of the rapidity distribution $\rho(t,k)$, for the exact numerical evolution (markers) and the analytic approximation (curve) at fixed $\gamma t\in\{0.5,1,2\}$ and two values of the dissipative rate $\gamma$.
(right) Distance (renormalized over the total length) from the exact state to the optimal Gaussian approximation $\delta(\hat{\rho},\hat{\rho}_G)$, and to the Gaussian state obtained from the analytical solution $\delta(\hat{\rho},\hat{\rho}_{GGE})$, for two different values of the total dissipative rate $\gamma = 0.5,\, 0.1$.} 
\label{fig:nk_gamma}
\end{figure*}

\section{Discussion on the accuracy of $t$-GGE} \label{A:Precision}

We described in Section \ref{sec:homogeneous} that the accuracy of rapidity distributions improves significantly in the presence of both gain and loss; this substantial influence even being of an order of magnitude between $n_s=0$ and $n_s=3/4$. Due to particle-hole symmetry, the same behavior is observed for gain only ($n_s=1$) and $n_s=1/4$. Precisely, for $n_s=0$ (resp. $1$), the loss (resp. gain) affects precision at its utmost. While with losses $t$-GGE tends to underestimate the rapidity profile for $k\sim0$, gains act oppositely for $k\sim\pi$. Given these observations, it seems reasonable to expect that accuracy would be highest when gain and loss are maximally balanced.
This is also suggested by the master equation, e.g., in the form of Eq. \eqref{eq:reduced Masteq}, where for $n_s=1/2$, the balancing between gain and loss term is maximized, and many terms cancel. 
As it can be seen in Figure \ref{fig:nk_gamma}, at $n_s=1/2$ the prediction of rapidity distribution is almost exact, and we immediately see the collpase onto the theoretical function of $\gamma t$. We can still see that the predicted $\rho_g(t,k)$ is not exact from the behaviour of $\delta(\hat\rho,\hat\rho_{GGE})$ versus $\delta(\hat\rho,\hat\rho_G)$, that do not overlap exactly at their maximum. However, great accuracy in the prediction of the rapidity distribution does not ultimately imply overall better precision for more complicated correlation functions, as they would be calculated on the $t$-GGE using Wick's theorem, while the actual state is non-Gaussian. It is interesting to notice that the best precision for the rapidity distribution is gotten in coincidence with maximal non-Gaussianity, so that the error on two-point functions and the correction to Wick's theorem will balance.

\bibliography{bib}

\begin{thebibliography}{71}%
\makeatletter
\providecommand \@ifxundefined [1]{%
 \@ifx{#1\undefined}
}%
\providecommand \@ifnum [1]{%
 \ifnum #1\expandafter \@firstoftwo
 \else \expandafter \@secondoftwo
 \fi
}%
\providecommand \@ifx [1]{%
 \ifx #1\expandafter \@firstoftwo
 \else \expandafter \@secondoftwo
 \fi
}%
\providecommand \natexlab [1]{#1}%
\providecommand \enquote  [1]{``#1''}%
\providecommand \bibnamefont  [1]{#1}%
\providecommand \bibfnamefont [1]{#1}%
\providecommand \citenamefont [1]{#1}%
\providecommand \href@noop [0]{\@secondoftwo}%
\providecommand \href [0]{\begingroup \@sanitize@url \@href}%
\providecommand \@href[1]{\@@startlink{#1}\@@href}%
\providecommand \@@href[1]{\endgroup#1\@@endlink}%
\providecommand \@sanitize@url [0]{\catcode `\\12\catcode `\$12\catcode `\&12\catcode `\#12\catcode `\^12\catcode `\_12\catcode `\%12\relax}%
\providecommand \@@startlink[1]{}%
\providecommand \@@endlink[0]{}%
\providecommand \url  [0]{\begingroup\@sanitize@url \@url }%
\providecommand \@url [1]{\endgroup\@href {#1}{\urlprefix }}%
\providecommand \urlprefix  [0]{URL }%
\providecommand \Eprint [0]{\href }%
\providecommand \doibase [0]{https://doi.org/}%
\providecommand \selectlanguage [0]{\@gobble}%
\providecommand \bibinfo  [0]{\@secondoftwo}%
\providecommand \bibfield  [0]{\@secondoftwo}%
\providecommand \translation [1]{[#1]}%
\providecommand \BibitemOpen [0]{}%
\providecommand \bibitemStop [0]{}%
\providecommand \bibitemNoStop [0]{.\EOS\space}%
\providecommand \EOS [0]{\spacefactor3000\relax}%
\providecommand \BibitemShut  [1]{\csname bibitem#1\endcsname}%
\let\auto@bib@innerbib\@empty
\bibitem [{\citenamefont {Rivas}\ and\ \citenamefont {Huelga}(2012)}]{rivas2012open}%
  \BibitemOpen
  \bibfield  {author} {\bibinfo {author} {\bibfnamefont {A.}~\bibnamefont {Rivas}}\ and\ \bibinfo {author} {\bibfnamefont {S.~F.}\ \bibnamefont {Huelga}},\ }\href@noop {} {\emph {\bibinfo {title} {Open quantum systems}}},\ Vol.~\bibinfo {volume} {10}\ (\bibinfo  {publisher} {Springer},\ \bibinfo {year} {2012})\BibitemShut {NoStop}%
\bibitem [{\citenamefont {Breuer}\ and\ \citenamefont {Petruccione}(2002)}]{breuer2002theory}%
  \BibitemOpen
  \bibfield  {author} {\bibinfo {author} {\bibfnamefont {H.-P.}\ \bibnamefont {Breuer}}\ and\ \bibinfo {author} {\bibfnamefont {F.}~\bibnamefont {Petruccione}},\ }\href@noop {} {\emph {\bibinfo {title} {The theory of open quantum systems}}}\ (\bibinfo  {publisher} {Oxford University Press, USA},\ \bibinfo {year} {2002})\BibitemShut {NoStop}%
\bibitem [{\citenamefont {Schlosshauer}(2004)}]{schlosshauer2004decoherence}%
  \BibitemOpen
  \bibfield  {author} {\bibinfo {author} {\bibfnamefont {M.}~\bibnamefont {Schlosshauer}},\ }\bibfield  {title} {\bibinfo {title} {Decoherence, the measurement problem, and interpretations of quantum mechanics},\ }\href@noop {} {\bibfield  {journal} {\bibinfo  {journal} {Reviews of Modern physics}\ }\textbf {\bibinfo {volume} {76}},\ \bibinfo {pages} {1267} (\bibinfo {year} {2004})}\BibitemShut {NoStop}%
\bibitem [{\citenamefont {Vinjanampathy}\ and\ \citenamefont {Anders}(2016)}]{vinjanampathy2016quantum}%
  \BibitemOpen
  \bibfield  {author} {\bibinfo {author} {\bibfnamefont {S.}~\bibnamefont {Vinjanampathy}}\ and\ \bibinfo {author} {\bibfnamefont {J.}~\bibnamefont {Anders}},\ }\bibfield  {title} {\bibinfo {title} {Quantum thermodynamics},\ }\href@noop {} {\bibfield  {journal} {\bibinfo  {journal} {Contemporary Physics}\ }\textbf {\bibinfo {volume} {57}},\ \bibinfo {pages} {545} (\bibinfo {year} {2016})}\BibitemShut {NoStop}%
\bibitem [{\citenamefont {Weiner}\ \emph {et~al.}(1999)\citenamefont {Weiner}, \citenamefont {Bagnato}, \citenamefont {Zilio},\ and\ \citenamefont {Julienne}}]{weiner1999experiments}%
  \BibitemOpen
  \bibfield  {author} {\bibinfo {author} {\bibfnamefont {J.}~\bibnamefont {Weiner}}, \bibinfo {author} {\bibfnamefont {V.~S.}\ \bibnamefont {Bagnato}}, \bibinfo {author} {\bibfnamefont {S.}~\bibnamefont {Zilio}},\ and\ \bibinfo {author} {\bibfnamefont {P.~S.}\ \bibnamefont {Julienne}},\ }\bibfield  {title} {\bibinfo {title} {Experiments and theory in cold and ultracold collisions},\ }\href@noop {} {\bibfield  {journal} {\bibinfo  {journal} {Reviews of Modern Physics}\ }\textbf {\bibinfo {volume} {71}},\ \bibinfo {pages} {1} (\bibinfo {year} {1999})}\BibitemShut {NoStop}%
\bibitem [{\citenamefont {Bouchoule}\ and\ \citenamefont {Dubail}(2022)}]{bouchoule2022generalized}%
  \BibitemOpen
  \bibfield  {author} {\bibinfo {author} {\bibfnamefont {I.}~\bibnamefont {Bouchoule}}\ and\ \bibinfo {author} {\bibfnamefont {J.}~\bibnamefont {Dubail}},\ }\bibfield  {title} {\bibinfo {title} {Generalized hydrodynamics in the one-dimensional bose gas: theory and experiments},\ }\href@noop {} {\bibfield  {journal} {\bibinfo  {journal} {Journal of Statistical Mechanics: Theory and Experiment}\ }\textbf {\bibinfo {volume} {2022}},\ \bibinfo {pages} {014003} (\bibinfo {year} {2022})}\BibitemShut {NoStop}%
\bibitem [{\citenamefont {Diehl}\ \emph {et~al.}(2008)\citenamefont {Diehl}, \citenamefont {Micheli}, \citenamefont {Kantian}, \citenamefont {Kraus}, \citenamefont {B{\"u}chler},\ and\ \citenamefont {Zoller}}]{diehl2008quantum}%
  \BibitemOpen
  \bibfield  {author} {\bibinfo {author} {\bibfnamefont {S.}~\bibnamefont {Diehl}}, \bibinfo {author} {\bibfnamefont {A.}~\bibnamefont {Micheli}}, \bibinfo {author} {\bibfnamefont {A.}~\bibnamefont {Kantian}}, \bibinfo {author} {\bibfnamefont {B.}~\bibnamefont {Kraus}}, \bibinfo {author} {\bibfnamefont {H.}~\bibnamefont {B{\"u}chler}},\ and\ \bibinfo {author} {\bibfnamefont {P.}~\bibnamefont {Zoller}},\ }\bibfield  {title} {\bibinfo {title} {Quantum states and phases in driven open quantum systems with cold atoms},\ }\href@noop {} {\bibfield  {journal} {\bibinfo  {journal} {Nature Physics}\ }\textbf {\bibinfo {volume} {4}},\ \bibinfo {pages} {878} (\bibinfo {year} {2008})}\BibitemShut {NoStop}%
\bibitem [{\citenamefont {Verstraete}\ \emph {et~al.}(2009)\citenamefont {Verstraete}, \citenamefont {Wolf},\ and\ \citenamefont {Ignacio~Cirac}}]{verstraete2009quantum}%
  \BibitemOpen
  \bibfield  {author} {\bibinfo {author} {\bibfnamefont {F.}~\bibnamefont {Verstraete}}, \bibinfo {author} {\bibfnamefont {M.~M.}\ \bibnamefont {Wolf}},\ and\ \bibinfo {author} {\bibfnamefont {J.}~\bibnamefont {Ignacio~Cirac}},\ }\bibfield  {title} {\bibinfo {title} {Quantum computation and quantum-state engineering driven by dissipation},\ }\href@noop {} {\bibfield  {journal} {\bibinfo  {journal} {Nature physics}\ }\textbf {\bibinfo {volume} {5}},\ \bibinfo {pages} {633} (\bibinfo {year} {2009})}\BibitemShut {NoStop}%
\bibitem [{\citenamefont {Harrington}\ \emph {et~al.}(2022)\citenamefont {Harrington}, \citenamefont {Mueller},\ and\ \citenamefont {Murch}}]{harrington2022engineered}%
  \BibitemOpen
  \bibfield  {author} {\bibinfo {author} {\bibfnamefont {P.~M.}\ \bibnamefont {Harrington}}, \bibinfo {author} {\bibfnamefont {E.~J.}\ \bibnamefont {Mueller}},\ and\ \bibinfo {author} {\bibfnamefont {K.~W.}\ \bibnamefont {Murch}},\ }\bibfield  {title} {\bibinfo {title} {Engineered dissipation for quantum information science},\ }\href@noop {} {\bibfield  {journal} {\bibinfo  {journal} {Nature Reviews Physics}\ }\textbf {\bibinfo {volume} {4}},\ \bibinfo {pages} {660} (\bibinfo {year} {2022})}\BibitemShut {NoStop}%
\bibitem [{\citenamefont {Cohen}\ and\ \citenamefont {Mirrahimi}(2014)}]{cohen2014dissipation}%
  \BibitemOpen
  \bibfield  {author} {\bibinfo {author} {\bibfnamefont {J.}~\bibnamefont {Cohen}}\ and\ \bibinfo {author} {\bibfnamefont {M.}~\bibnamefont {Mirrahimi}},\ }\bibfield  {title} {\bibinfo {title} {Dissipation-induced continuous quantum error correction for superconducting circuits},\ }\href@noop {} {\bibfield  {journal} {\bibinfo  {journal} {Physical Review A}\ }\textbf {\bibinfo {volume} {90}},\ \bibinfo {pages} {062344} (\bibinfo {year} {2014})}\BibitemShut {NoStop}%
\bibitem [{\citenamefont {Lebreuilly}\ \emph {et~al.}(2021)\citenamefont {Lebreuilly}, \citenamefont {Noh}, \citenamefont {Wang}, \citenamefont {Girvin},\ and\ \citenamefont {Jiang}}]{lebreuilly2021autonomous}%
  \BibitemOpen
  \bibfield  {author} {\bibinfo {author} {\bibfnamefont {J.}~\bibnamefont {Lebreuilly}}, \bibinfo {author} {\bibfnamefont {K.}~\bibnamefont {Noh}}, \bibinfo {author} {\bibfnamefont {C.-H.}\ \bibnamefont {Wang}}, \bibinfo {author} {\bibfnamefont {S.~M.}\ \bibnamefont {Girvin}},\ and\ \bibinfo {author} {\bibfnamefont {L.}~\bibnamefont {Jiang}},\ }\bibfield  {title} {\bibinfo {title} {Autonomous quantum error correction and quantum computation},\ }\href@noop {} {\bibfield  {journal} {\bibinfo  {journal} {arXiv preprint arXiv:2103.05007}\ } (\bibinfo {year} {2021})}\BibitemShut {NoStop}%
\bibitem [{\citenamefont {Reiter}\ \emph {et~al.}(2017)\citenamefont {Reiter}, \citenamefont {S{\o}rensen}, \citenamefont {Zoller},\ and\ \citenamefont {Muschik}}]{reiter2017dissipative}%
  \BibitemOpen
  \bibfield  {author} {\bibinfo {author} {\bibfnamefont {F.}~\bibnamefont {Reiter}}, \bibinfo {author} {\bibfnamefont {A.~S.}\ \bibnamefont {S{\o}rensen}}, \bibinfo {author} {\bibfnamefont {P.}~\bibnamefont {Zoller}},\ and\ \bibinfo {author} {\bibfnamefont {C.}~\bibnamefont {Muschik}},\ }\bibfield  {title} {\bibinfo {title} {Dissipative quantum error correction and application to quantum sensing with trapped ions},\ }\href@noop {} {\bibfield  {journal} {\bibinfo  {journal} {Nature communications}\ }\textbf {\bibinfo {volume} {8}},\ \bibinfo {pages} {1822} (\bibinfo {year} {2017})}\BibitemShut {NoStop}%
\bibitem [{\citenamefont {Str{\"u}mpfer}\ and\ \citenamefont {Schulten}(2012)}]{strümpfer2012open}%
  \BibitemOpen
  \bibfield  {author} {\bibinfo {author} {\bibfnamefont {J.}~\bibnamefont {Str{\"u}mpfer}}\ and\ \bibinfo {author} {\bibfnamefont {K.}~\bibnamefont {Schulten}},\ }\bibfield  {title} {\bibinfo {title} {Open quantum dynamics calculations with the hierarchy equations of motion on parallel computers},\ }\href@noop {} {\bibfield  {journal} {\bibinfo  {journal} {Journal of chemical theory and computation}\ }\textbf {\bibinfo {volume} {8}},\ \bibinfo {pages} {2808} (\bibinfo {year} {2012})}\BibitemShut {NoStop}%
\bibitem [{\citenamefont {Nitzan}(2024)}]{nitzan2024chemical}%
  \BibitemOpen
  \bibfield  {author} {\bibinfo {author} {\bibfnamefont {A.}~\bibnamefont {Nitzan}},\ }\href@noop {} {\emph {\bibinfo {title} {Chemical dynamics in condensed phases: relaxation, transfer, and reactions in condensed molecular systems}}}\ (\bibinfo  {publisher} {Oxford university press},\ \bibinfo {year} {2024})\BibitemShut {NoStop}%
\bibitem [{\citenamefont {May}\ and\ \citenamefont {K{\"u}hn}(2023)}]{may2023charge}%
  \BibitemOpen
  \bibfield  {author} {\bibinfo {author} {\bibfnamefont {V.}~\bibnamefont {May}}\ and\ \bibinfo {author} {\bibfnamefont {O.}~\bibnamefont {K{\"u}hn}},\ }\href@noop {} {\emph {\bibinfo {title} {Charge and energy transfer dynamics in molecular systems}}}\ (\bibinfo  {publisher} {John Wiley \& Sons},\ \bibinfo {year} {2023})\BibitemShut {NoStop}%
\bibitem [{\citenamefont {Schlimgen}\ \emph {et~al.}(2021)\citenamefont {Schlimgen}, \citenamefont {Head-Marsden}, \citenamefont {Sager}, \citenamefont {Narang},\ and\ \citenamefont {Mazziotti}}]{schlimgen2021quantum}%
  \BibitemOpen
  \bibfield  {author} {\bibinfo {author} {\bibfnamefont {A.~W.}\ \bibnamefont {Schlimgen}}, \bibinfo {author} {\bibfnamefont {K.}~\bibnamefont {Head-Marsden}}, \bibinfo {author} {\bibfnamefont {L.~M.}\ \bibnamefont {Sager}}, \bibinfo {author} {\bibfnamefont {P.}~\bibnamefont {Narang}},\ and\ \bibinfo {author} {\bibfnamefont {D.~A.}\ \bibnamefont {Mazziotti}},\ }\bibfield  {title} {\bibinfo {title} {Quantum simulation of open quantum systems using a unitary decomposition of operators},\ }\href@noop {} {\bibfield  {journal} {\bibinfo  {journal} {Physical Review Letters}\ }\textbf {\bibinfo {volume} {127}},\ \bibinfo {pages} {270503} (\bibinfo {year} {2021})}\BibitemShut {NoStop}%
\bibitem [{\citenamefont {Blankenship}(2021)}]{blankenship2021molecular}%
  \BibitemOpen
  \bibfield  {author} {\bibinfo {author} {\bibfnamefont {R.~E.}\ \bibnamefont {Blankenship}},\ }\href@noop {} {\emph {\bibinfo {title} {Molecular mechanisms of photosynthesis}}}\ (\bibinfo  {publisher} {John Wiley \& Sons},\ \bibinfo {year} {2021})\BibitemShut {NoStop}%
\bibitem [{\citenamefont {Dalibard}\ \emph {et~al.}(1992)\citenamefont {Dalibard}, \citenamefont {Castin},\ and\ \citenamefont {M{\o}lmer}}]{dalibard1992wave}%
  \BibitemOpen
  \bibfield  {author} {\bibinfo {author} {\bibfnamefont {J.}~\bibnamefont {Dalibard}}, \bibinfo {author} {\bibfnamefont {Y.}~\bibnamefont {Castin}},\ and\ \bibinfo {author} {\bibfnamefont {K.}~\bibnamefont {M{\o}lmer}},\ }\bibfield  {title} {\bibinfo {title} {Wave-function approach to dissipative processes in quantum optics},\ }\href@noop {} {\bibfield  {journal} {\bibinfo  {journal} {Physical review letters}\ }\textbf {\bibinfo {volume} {68}},\ \bibinfo {pages} {580} (\bibinfo {year} {1992})}\BibitemShut {NoStop}%
\bibitem [{\citenamefont {Daley}(2014)}]{daley2014quantum}%
  \BibitemOpen
  \bibfield  {author} {\bibinfo {author} {\bibfnamefont {A.~J.}\ \bibnamefont {Daley}},\ }\bibfield  {title} {\bibinfo {title} {Quantum trajectories and open many-body quantum systems},\ }\href@noop {} {\bibfield  {journal} {\bibinfo  {journal} {Advances in Physics}\ }\textbf {\bibinfo {volume} {63}},\ \bibinfo {pages} {77} (\bibinfo {year} {2014})}\BibitemShut {NoStop}%
\bibitem [{\citenamefont {Verstraete}\ \emph {et~al.}(2004)\citenamefont {Verstraete}, \citenamefont {Garcia-Ripoll},\ and\ \citenamefont {Cirac}}]{verstraete2004matrix}%
  \BibitemOpen
  \bibfield  {author} {\bibinfo {author} {\bibfnamefont {F.}~\bibnamefont {Verstraete}}, \bibinfo {author} {\bibfnamefont {J.~J.}\ \bibnamefont {Garcia-Ripoll}},\ and\ \bibinfo {author} {\bibfnamefont {J.~I.}\ \bibnamefont {Cirac}},\ }\bibfield  {title} {\bibinfo {title} {Matrix product density operators: Simulation of finite-temperature and dissipative systems},\ }\href@noop {} {\bibfield  {journal} {\bibinfo  {journal} {Physical review letters}\ }\textbf {\bibinfo {volume} {93}},\ \bibinfo {pages} {207204} (\bibinfo {year} {2004})}\BibitemShut {NoStop}%
\bibitem [{\citenamefont {Zwolak}\ and\ \citenamefont {Vidal}(2004)}]{zwolak2004mixed}%
  \BibitemOpen
  \bibfield  {author} {\bibinfo {author} {\bibfnamefont {M.}~\bibnamefont {Zwolak}}\ and\ \bibinfo {author} {\bibfnamefont {G.}~\bibnamefont {Vidal}},\ }\bibfield  {title} {\bibinfo {title} {Mixed-state dynamics in one-dimensional quantum lattice systems:<? format?> a time-dependent superoperator renormalization algorithm},\ }\href@noop {} {\bibfield  {journal} {\bibinfo  {journal} {Physical review letters}\ }\textbf {\bibinfo {volume} {93}},\ \bibinfo {pages} {207205} (\bibinfo {year} {2004})}\BibitemShut {NoStop}%
\bibitem [{\citenamefont {Verstraete}\ \emph {et~al.}(2008)\citenamefont {Verstraete}, \citenamefont {Murg},\ and\ \citenamefont {Cirac}}]{verstraete2008matrix}%
  \BibitemOpen
  \bibfield  {author} {\bibinfo {author} {\bibfnamefont {F.}~\bibnamefont {Verstraete}}, \bibinfo {author} {\bibfnamefont {V.}~\bibnamefont {Murg}},\ and\ \bibinfo {author} {\bibfnamefont {J.~I.}\ \bibnamefont {Cirac}},\ }\bibfield  {title} {\bibinfo {title} {Matrix product states, projected entangled pair states, and variational renormalization group methods for quantum spin systems},\ }\href@noop {} {\bibfield  {journal} {\bibinfo  {journal} {Advances in physics}\ }\textbf {\bibinfo {volume} {57}},\ \bibinfo {pages} {143} (\bibinfo {year} {2008})}\BibitemShut {NoStop}%
\bibitem [{\citenamefont {Hartmann}\ and\ \citenamefont {Carleo}(2019)}]{hartmann2019neural}%
  \BibitemOpen
  \bibfield  {author} {\bibinfo {author} {\bibfnamefont {M.~J.}\ \bibnamefont {Hartmann}}\ and\ \bibinfo {author} {\bibfnamefont {G.}~\bibnamefont {Carleo}},\ }\bibfield  {title} {\bibinfo {title} {Neural-network approach to dissipative quantum many-body dynamics},\ }\href@noop {} {\bibfield  {journal} {\bibinfo  {journal} {Physical review letters}\ }\textbf {\bibinfo {volume} {122}},\ \bibinfo {pages} {250502} (\bibinfo {year} {2019})}\BibitemShut {NoStop}%
\bibitem [{\citenamefont {Weimer}\ \emph {et~al.}(2021)\citenamefont {Weimer}, \citenamefont {Kshetrimayum},\ and\ \citenamefont {Or{\'u}s}}]{weimer2021simulation}%
  \BibitemOpen
  \bibfield  {author} {\bibinfo {author} {\bibfnamefont {H.}~\bibnamefont {Weimer}}, \bibinfo {author} {\bibfnamefont {A.}~\bibnamefont {Kshetrimayum}},\ and\ \bibinfo {author} {\bibfnamefont {R.}~\bibnamefont {Or{\'u}s}},\ }\bibfield  {title} {\bibinfo {title} {Simulation methods for open quantum many-body systems},\ }\href@noop {} {\bibfield  {journal} {\bibinfo  {journal} {Reviews of Modern Physics}\ }\textbf {\bibinfo {volume} {93}},\ \bibinfo {pages} {015008} (\bibinfo {year} {2021})}\BibitemShut {NoStop}%
\bibitem [{\citenamefont {Lange}\ \emph {et~al.}(2017)\citenamefont {Lange}, \citenamefont {Lenar{\v{c}}i{\v{c}}},\ and\ \citenamefont {Rosch}}]{lange2017pumping}%
  \BibitemOpen
  \bibfield  {author} {\bibinfo {author} {\bibfnamefont {F.}~\bibnamefont {Lange}}, \bibinfo {author} {\bibfnamefont {Z.}~\bibnamefont {Lenar{\v{c}}i{\v{c}}}},\ and\ \bibinfo {author} {\bibfnamefont {A.}~\bibnamefont {Rosch}},\ }\bibfield  {title} {\bibinfo {title} {Pumping approximately integrable systems},\ }\href@noop {} {\bibfield  {journal} {\bibinfo  {journal} {Nature communications}\ }\textbf {\bibinfo {volume} {8}},\ \bibinfo {pages} {15767} (\bibinfo {year} {2017})}\BibitemShut {NoStop}%
\bibitem [{\citenamefont {Lange}\ \emph {et~al.}(2018)\citenamefont {Lange}, \citenamefont {Lenar{\v{c}}i{\v{c}}},\ and\ \citenamefont {Rosch}}]{lange2018time}%
  \BibitemOpen
  \bibfield  {author} {\bibinfo {author} {\bibfnamefont {F.}~\bibnamefont {Lange}}, \bibinfo {author} {\bibfnamefont {Z.}~\bibnamefont {Lenar{\v{c}}i{\v{c}}}},\ and\ \bibinfo {author} {\bibfnamefont {A.}~\bibnamefont {Rosch}},\ }\bibfield  {title} {\bibinfo {title} {Time-dependent generalized gibbs ensembles in open quantum systems},\ }\href@noop {} {\bibfield  {journal} {\bibinfo  {journal} {Physical Review B}\ }\textbf {\bibinfo {volume} {97}},\ \bibinfo {pages} {165138} (\bibinfo {year} {2018})}\BibitemShut {NoStop}%
\bibitem [{\citenamefont {Lenar{\v{c}}i{\v{c}}}\ \emph {et~al.}(2018)\citenamefont {Lenar{\v{c}}i{\v{c}}}, \citenamefont {Lange},\ and\ \citenamefont {Rosch}}]{lenarvcivc2018perturbative}%
  \BibitemOpen
  \bibfield  {author} {\bibinfo {author} {\bibfnamefont {Z.}~\bibnamefont {Lenar{\v{c}}i{\v{c}}}}, \bibinfo {author} {\bibfnamefont {F.}~\bibnamefont {Lange}},\ and\ \bibinfo {author} {\bibfnamefont {A.}~\bibnamefont {Rosch}},\ }\bibfield  {title} {\bibinfo {title} {Perturbative approach to weakly driven many-particle systems in the presence of approximate conservation laws},\ }\href@noop {} {\bibfield  {journal} {\bibinfo  {journal} {Physical Review B}\ }\textbf {\bibinfo {volume} {97}},\ \bibinfo {pages} {024302} (\bibinfo {year} {2018})}\BibitemShut {NoStop}%
\bibitem [{\citenamefont {Bouchoule}\ \emph {et~al.}(2020)\citenamefont {Bouchoule}, \citenamefont {Doyon},\ and\ \citenamefont {Dubail}}]{bouchoule2020effect}%
  \BibitemOpen
  \bibfield  {author} {\bibinfo {author} {\bibfnamefont {I.}~\bibnamefont {Bouchoule}}, \bibinfo {author} {\bibfnamefont {B.}~\bibnamefont {Doyon}},\ and\ \bibinfo {author} {\bibfnamefont {J.}~\bibnamefont {Dubail}},\ }\bibfield  {title} {\bibinfo {title} {The effect of atom losses on the distribution of rapidities in the one-dimensional bose gas},\ }\href@noop {} {\bibfield  {journal} {\bibinfo  {journal} {SciPost Physics}\ }\textbf {\bibinfo {volume} {9}},\ \bibinfo {pages} {044} (\bibinfo {year} {2020})}\BibitemShut {NoStop}%
\bibitem [{\citenamefont {Bouchoule}\ and\ \citenamefont {Dubail}(2021)}]{bouchoule2021breakdown}%
  \BibitemOpen
  \bibfield  {author} {\bibinfo {author} {\bibfnamefont {I.}~\bibnamefont {Bouchoule}}\ and\ \bibinfo {author} {\bibfnamefont {J.}~\bibnamefont {Dubail}},\ }\bibfield  {title} {\bibinfo {title} {Breakdown of tan’s relation in lossy one-dimensional bose gases},\ }\href@noop {} {\bibfield  {journal} {\bibinfo  {journal} {Physical Review Letters}\ }\textbf {\bibinfo {volume} {126}},\ \bibinfo {pages} {160603} (\bibinfo {year} {2021})}\BibitemShut {NoStop}%
\bibitem [{\citenamefont {Rossini}\ \emph {et~al.}(2021)\citenamefont {Rossini}, \citenamefont {Ghermaoui}, \citenamefont {Aguilera}, \citenamefont {Vatr{\'e}}, \citenamefont {Bouganne}, \citenamefont {Beugnon}, \citenamefont {Gerbier},\ and\ \citenamefont {Mazza}}]{rossini2021strong}%
  \BibitemOpen
  \bibfield  {author} {\bibinfo {author} {\bibfnamefont {D.}~\bibnamefont {Rossini}}, \bibinfo {author} {\bibfnamefont {A.}~\bibnamefont {Ghermaoui}}, \bibinfo {author} {\bibfnamefont {M.~B.}\ \bibnamefont {Aguilera}}, \bibinfo {author} {\bibfnamefont {R.}~\bibnamefont {Vatr{\'e}}}, \bibinfo {author} {\bibfnamefont {R.}~\bibnamefont {Bouganne}}, \bibinfo {author} {\bibfnamefont {J.}~\bibnamefont {Beugnon}}, \bibinfo {author} {\bibfnamefont {F.}~\bibnamefont {Gerbier}},\ and\ \bibinfo {author} {\bibfnamefont {L.}~\bibnamefont {Mazza}},\ }\bibfield  {title} {\bibinfo {title} {Strong correlations in lossy one-dimensional quantum gases: From the quantum zeno effect to the generalized gibbs ensemble},\ }\href@noop {} {\bibfield  {journal} {\bibinfo  {journal} {Physical Review A}\ }\textbf {\bibinfo {volume} {103}},\ \bibinfo {pages} {L060201} (\bibinfo {year} {2021})}\BibitemShut {NoStop}%
\bibitem [{\citenamefont {Rosso}\ \emph {et~al.}(2021)\citenamefont {Rosso}, \citenamefont {Rossini}, \citenamefont {Biella},\ and\ \citenamefont {Mazza}}]{rosso2021one}%
  \BibitemOpen
  \bibfield  {author} {\bibinfo {author} {\bibfnamefont {L.}~\bibnamefont {Rosso}}, \bibinfo {author} {\bibfnamefont {D.}~\bibnamefont {Rossini}}, \bibinfo {author} {\bibfnamefont {A.}~\bibnamefont {Biella}},\ and\ \bibinfo {author} {\bibfnamefont {L.}~\bibnamefont {Mazza}},\ }\bibfield  {title} {\bibinfo {title} {One-dimensional spin-1/2 fermionic gases with two-body losses: Weak dissipation and spin conservation},\ }\href@noop {} {\bibfield  {journal} {\bibinfo  {journal} {Physical Review A}\ }\textbf {\bibinfo {volume} {104}},\ \bibinfo {pages} {053305} (\bibinfo {year} {2021})}\BibitemShut {NoStop}%
\bibitem [{\citenamefont {Reiter}\ \emph {et~al.}(2021)\citenamefont {Reiter}, \citenamefont {Lange}, \citenamefont {Jain}, \citenamefont {Grau}, \citenamefont {Home},\ and\ \citenamefont {Lenar{\v{c}}i{\v{c}}}}]{reiter2021engineering}%
  \BibitemOpen
  \bibfield  {author} {\bibinfo {author} {\bibfnamefont {F.}~\bibnamefont {Reiter}}, \bibinfo {author} {\bibfnamefont {F.}~\bibnamefont {Lange}}, \bibinfo {author} {\bibfnamefont {S.}~\bibnamefont {Jain}}, \bibinfo {author} {\bibfnamefont {M.}~\bibnamefont {Grau}}, \bibinfo {author} {\bibfnamefont {J.~P.}\ \bibnamefont {Home}},\ and\ \bibinfo {author} {\bibfnamefont {Z.}~\bibnamefont {Lenar{\v{c}}i{\v{c}}}},\ }\bibfield  {title} {\bibinfo {title} {Engineering generalized gibbs ensembles with trapped ions},\ }\href@noop {} {\bibfield  {journal} {\bibinfo  {journal} {Physical Review Research}\ }\textbf {\bibinfo {volume} {3}},\ \bibinfo {pages} {033142} (\bibinfo {year} {2021})}\BibitemShut {NoStop}%
\bibitem [{\citenamefont {Rosso}\ \emph {et~al.}(2022)\citenamefont {Rosso}, \citenamefont {Biella},\ and\ \citenamefont {Mazza}}]{rosso2022one}%
  \BibitemOpen
  \bibfield  {author} {\bibinfo {author} {\bibfnamefont {L.}~\bibnamefont {Rosso}}, \bibinfo {author} {\bibfnamefont {A.}~\bibnamefont {Biella}},\ and\ \bibinfo {author} {\bibfnamefont {L.}~\bibnamefont {Mazza}},\ }\bibfield  {title} {\bibinfo {title} {The one-dimensional bose gas with strong two-body losses: the effect of the harmonic confinement},\ }\href@noop {} {\bibfield  {journal} {\bibinfo  {journal} {SciPost Physics}\ }\textbf {\bibinfo {volume} {12}},\ \bibinfo {pages} {044} (\bibinfo {year} {2022})}\BibitemShut {NoStop}%
\bibitem [{\citenamefont {Riggio}\ \emph {et~al.}(2024)\citenamefont {Riggio}, \citenamefont {Rosso}, \citenamefont {Karevski},\ and\ \citenamefont {Dubail}}]{riggio2023effects}%
  \BibitemOpen
  \bibfield  {author} {\bibinfo {author} {\bibfnamefont {F.}~\bibnamefont {Riggio}}, \bibinfo {author} {\bibfnamefont {L.}~\bibnamefont {Rosso}}, \bibinfo {author} {\bibfnamefont {D.}~\bibnamefont {Karevski}},\ and\ \bibinfo {author} {\bibfnamefont {J.}~\bibnamefont {Dubail}},\ }\bibfield  {title} {\bibinfo {title} {Effects of atom losses on a one-dimensional lattice gas of hard-core bosons},\ }\href@noop {} {\bibfield  {journal} {\bibinfo  {journal} {Physical Review A}\ }\textbf {\bibinfo {volume} {109}},\ \bibinfo {pages} {023311} (\bibinfo {year} {2024})}\BibitemShut {NoStop}%
\bibitem [{\citenamefont {Maki}\ \emph {et~al.}(2024)\citenamefont {Maki}, \citenamefont {Rosso}, \citenamefont {Mazza},\ and\ \citenamefont {Biella}}]{maki2024loss}%
  \BibitemOpen
  \bibfield  {author} {\bibinfo {author} {\bibfnamefont {J.}~\bibnamefont {Maki}}, \bibinfo {author} {\bibfnamefont {L.}~\bibnamefont {Rosso}}, \bibinfo {author} {\bibfnamefont {L.}~\bibnamefont {Mazza}},\ and\ \bibinfo {author} {\bibfnamefont {A.}~\bibnamefont {Biella}},\ }\bibfield  {title} {\bibinfo {title} {Loss induced collective mode in one-dimensional bose gases},\ }\href@noop {} {\bibfield  {journal} {\bibinfo  {journal} {arXiv preprint arXiv:2402.05824}\ } (\bibinfo {year} {2024})}\BibitemShut {NoStop}%
\bibitem [{\citenamefont {Gerbino}\ \emph {et~al.}(2024{\natexlab{a}})\citenamefont {Gerbino}, \citenamefont {Lesanovsky},\ and\ \citenamefont {Perfetto}}]{gerbino2024large}%
  \BibitemOpen
  \bibfield  {author} {\bibinfo {author} {\bibfnamefont {F.}~\bibnamefont {Gerbino}}, \bibinfo {author} {\bibfnamefont {I.}~\bibnamefont {Lesanovsky}},\ and\ \bibinfo {author} {\bibfnamefont {G.}~\bibnamefont {Perfetto}},\ }\bibfield  {title} {\bibinfo {title} {Large-scale universality in quantum reaction-diffusion from keldysh field theory},\ }\href@noop {} {\bibfield  {journal} {\bibinfo  {journal} {Physical Review B}\ }\textbf {\bibinfo {volume} {109}},\ \bibinfo {pages} {L220304} (\bibinfo {year} {2024}{\natexlab{a}})}\BibitemShut {NoStop}%
\bibitem [{\citenamefont {Gerbino}\ \emph {et~al.}(2024{\natexlab{b}})\citenamefont {Gerbino}, \citenamefont {Lesanovsky},\ and\ \citenamefont {Perfetto}}]{gerbino2024kinetics}%
  \BibitemOpen
  \bibfield  {author} {\bibinfo {author} {\bibfnamefont {F.}~\bibnamefont {Gerbino}}, \bibinfo {author} {\bibfnamefont {I.}~\bibnamefont {Lesanovsky}},\ and\ \bibinfo {author} {\bibfnamefont {G.}~\bibnamefont {Perfetto}},\ }\bibfield  {title} {\bibinfo {title} {Kinetics of quantum reaction-diffusion systems},\ }\href@noop {} {\bibfield  {journal} {\bibinfo  {journal} {arXiv preprint arXiv:2406.20028}\ } (\bibinfo {year} {2024}{\natexlab{b}})}\BibitemShut {NoStop}%
\bibitem [{\citenamefont {Rowlands}\ \emph {et~al.}(2024)\citenamefont {Rowlands}, \citenamefont {Lesanovsky},\ and\ \citenamefont {Perfetto}}]{rowlands2024quantum}%
  \BibitemOpen
  \bibfield  {author} {\bibinfo {author} {\bibfnamefont {S.}~\bibnamefont {Rowlands}}, \bibinfo {author} {\bibfnamefont {I.}~\bibnamefont {Lesanovsky}},\ and\ \bibinfo {author} {\bibfnamefont {G.}~\bibnamefont {Perfetto}},\ }\bibfield  {title} {\bibinfo {title} {Quantum reaction-limited reaction--diffusion dynamics of noninteracting bose gases},\ }\href@noop {} {\bibfield  {journal} {\bibinfo  {journal} {New Journal of Physics}\ }\textbf {\bibinfo {volume} {26}},\ \bibinfo {pages} {043010} (\bibinfo {year} {2024})}\BibitemShut {NoStop}%
\bibitem [{\citenamefont {Z{\"u}ndel}\ \emph {et~al.}(2024)\citenamefont {Z{\"u}ndel}, \citenamefont {Mazza}, \citenamefont {Canet},\ and\ \citenamefont {Minguzzi}}]{zundel2024space}%
  \BibitemOpen
  \bibfield  {author} {\bibinfo {author} {\bibfnamefont {M.}~\bibnamefont {Z{\"u}ndel}}, \bibinfo {author} {\bibfnamefont {L.}~\bibnamefont {Mazza}}, \bibinfo {author} {\bibfnamefont {L.}~\bibnamefont {Canet}},\ and\ \bibinfo {author} {\bibfnamefont {A.}~\bibnamefont {Minguzzi}},\ }\bibfield  {title} {\bibinfo {title} {Space-time first-order correlations of an open bose-hubbard model with incoherent pump and loss},\ }\href@noop {} {\bibfield  {journal} {\bibinfo  {journal} {arXiv preprint arXiv:2405.19972}\ } (\bibinfo {year} {2024})}\BibitemShut {NoStop}%
\bibitem [{\citenamefont {Ul{\v{c}}akar}\ and\ \citenamefont {Lenar{\v{c}}i{\v{c}}}(2024)}]{ulvcakar2024generalized}%
  \BibitemOpen
  \bibfield  {author} {\bibinfo {author} {\bibfnamefont {I.}~\bibnamefont {Ul{\v{c}}akar}}\ and\ \bibinfo {author} {\bibfnamefont {Z.}~\bibnamefont {Lenar{\v{c}}i{\v{c}}}},\ }\bibfield  {title} {\bibinfo {title} {Generalized gibbs ensembles in weakly interacting dissipative systems and digital quantum computers},\ }\href@noop {} {\bibfield  {journal} {\bibinfo  {journal} {arXiv preprint arXiv:2406.17033}\ } (\bibinfo {year} {2024})}\BibitemShut {NoStop}%
\bibitem [{\citenamefont {Perfetto}\ \emph {et~al.}(2023{\natexlab{a}})\citenamefont {Perfetto}, \citenamefont {Carollo}, \citenamefont {Garrahan},\ and\ \citenamefont {Lesanovsky}}]{Perfetto_PRL2023}%
  \BibitemOpen
  \bibfield  {author} {\bibinfo {author} {\bibfnamefont {G.}~\bibnamefont {Perfetto}}, \bibinfo {author} {\bibfnamefont {F.}~\bibnamefont {Carollo}}, \bibinfo {author} {\bibfnamefont {J.~P.}\ \bibnamefont {Garrahan}},\ and\ \bibinfo {author} {\bibfnamefont {I.}~\bibnamefont {Lesanovsky}},\ }\bibfield  {title} {\bibinfo {title} {Reaction-limited quantum reaction-diffusion dynamics},\ }\href@noop {} {\bibfield  {journal} {\bibinfo  {journal} {Phys. Rev. Lett.}\ }\textbf {\bibinfo {volume} {130}},\ \bibinfo {pages} {210402} (\bibinfo {year} {2023}{\natexlab{a}})}\BibitemShut {NoStop}%
\bibitem [{\citenamefont {Perfetto}\ \emph {et~al.}(2023{\natexlab{b}})\citenamefont {Perfetto}, \citenamefont {Carollo}, \citenamefont {Garrahan},\ and\ \citenamefont {Lesanovsky}}]{Perfetto_PRE2023}%
  \BibitemOpen
  \bibfield  {author} {\bibinfo {author} {\bibfnamefont {G.}~\bibnamefont {Perfetto}}, \bibinfo {author} {\bibfnamefont {F.}~\bibnamefont {Carollo}}, \bibinfo {author} {\bibfnamefont {J.~P.}\ \bibnamefont {Garrahan}},\ and\ \bibinfo {author} {\bibfnamefont {I.}~\bibnamefont {Lesanovsky}},\ }\bibfield  {title} {\bibinfo {title} {Quantum reaction-limited reaction-diffusion dynamics of annihilation processes},\ }\href@noop {} {\bibfield  {journal} {\bibinfo  {journal} {Phys. Rev. E}\ }\textbf {\bibinfo {volume} {108}},\ \bibinfo {pages} {064104} (\bibinfo {year} {2023}{\natexlab{b}})}\BibitemShut {NoStop}%
\bibitem [{\citenamefont {Rigol}\ \emph {et~al.}(2007)\citenamefont {Rigol}, \citenamefont {Dunjko}, \citenamefont {Yurovsky},\ and\ \citenamefont {Olshanii}}]{rigol2007relaxation}%
  \BibitemOpen
  \bibfield  {author} {\bibinfo {author} {\bibfnamefont {M.}~\bibnamefont {Rigol}}, \bibinfo {author} {\bibfnamefont {V.}~\bibnamefont {Dunjko}}, \bibinfo {author} {\bibfnamefont {V.}~\bibnamefont {Yurovsky}},\ and\ \bibinfo {author} {\bibfnamefont {M.}~\bibnamefont {Olshanii}},\ }\bibfield  {title} {\bibinfo {title} {Relaxation in a completely integrable many-body quantum system: an ab initio study of the dynamics of the highly excited states of 1d lattice hard-core bosons},\ }\href@noop {} {\bibfield  {journal} {\bibinfo  {journal} {Physical review letters}\ }\textbf {\bibinfo {volume} {98}},\ \bibinfo {pages} {050405} (\bibinfo {year} {2007})}\BibitemShut {NoStop}%
\bibitem [{\citenamefont {Rigol}\ \emph {et~al.}(2008)\citenamefont {Rigol}, \citenamefont {Dunjko},\ and\ \citenamefont {Olshanii}}]{rigol2008thermalization}%
  \BibitemOpen
  \bibfield  {author} {\bibinfo {author} {\bibfnamefont {M.}~\bibnamefont {Rigol}}, \bibinfo {author} {\bibfnamefont {V.}~\bibnamefont {Dunjko}},\ and\ \bibinfo {author} {\bibfnamefont {M.}~\bibnamefont {Olshanii}},\ }\bibfield  {title} {\bibinfo {title} {Thermalization and its mechanism for generic isolated quantum systems},\ }\href@noop {} {\bibfield  {journal} {\bibinfo  {journal} {Nature}\ }\textbf {\bibinfo {volume} {452}},\ \bibinfo {pages} {854} (\bibinfo {year} {2008})}\BibitemShut {NoStop}%
\bibitem [{\citenamefont {Vidmar}\ and\ \citenamefont {Rigol}(2016)}]{vidmar2016generalized}%
  \BibitemOpen
  \bibfield  {author} {\bibinfo {author} {\bibfnamefont {L.}~\bibnamefont {Vidmar}}\ and\ \bibinfo {author} {\bibfnamefont {M.}~\bibnamefont {Rigol}},\ }\bibfield  {title} {\bibinfo {title} {Generalized gibbs ensemble in integrable lattice models},\ }\href@noop {} {\bibfield  {journal} {\bibinfo  {journal} {Journal of Statistical Mechanics: Theory and Experiment}\ }\textbf {\bibinfo {volume} {2016}},\ \bibinfo {pages} {064007} (\bibinfo {year} {2016})}\BibitemShut {NoStop}%
\bibitem [{\citenamefont {Essler}\ and\ \citenamefont {Fagotti}(2016)}]{essler2016quench}%
  \BibitemOpen
  \bibfield  {author} {\bibinfo {author} {\bibfnamefont {F.~H.}\ \bibnamefont {Essler}}\ and\ \bibinfo {author} {\bibfnamefont {M.}~\bibnamefont {Fagotti}},\ }\bibfield  {title} {\bibinfo {title} {Quench dynamics and relaxation in isolated integrable quantum spin chains},\ }\href@noop {} {\bibfield  {journal} {\bibinfo  {journal} {Journal of Statistical Mechanics: Theory and Experiment}\ }\textbf {\bibinfo {volume} {2016}},\ \bibinfo {pages} {064002} (\bibinfo {year} {2016})}\BibitemShut {NoStop}%
\bibitem [{\citenamefont {Fagotti}\ and\ \citenamefont {Essler}(2013)}]{fagotti2013reduced}%
  \BibitemOpen
  \bibfield  {author} {\bibinfo {author} {\bibfnamefont {M.}~\bibnamefont {Fagotti}}\ and\ \bibinfo {author} {\bibfnamefont {F.~H.}\ \bibnamefont {Essler}},\ }\bibfield  {title} {\bibinfo {title} {Reduced density matrix after a quantum quench},\ }\href@noop {} {\bibfield  {journal} {\bibinfo  {journal} {Physical Review B—Condensed Matter and Materials Physics}\ }\textbf {\bibinfo {volume} {87}},\ \bibinfo {pages} {245107} (\bibinfo {year} {2013})}\BibitemShut {NoStop}%
\bibitem [{\citenamefont {Lux}\ \emph {et~al.}(2014)\citenamefont {Lux}, \citenamefont {M{\"u}ller}, \citenamefont {Mitra},\ and\ \citenamefont {Rosch}}]{lux2014hydrodynamic}%
  \BibitemOpen
  \bibfield  {author} {\bibinfo {author} {\bibfnamefont {J.}~\bibnamefont {Lux}}, \bibinfo {author} {\bibfnamefont {J.}~\bibnamefont {M{\"u}ller}}, \bibinfo {author} {\bibfnamefont {A.}~\bibnamefont {Mitra}},\ and\ \bibinfo {author} {\bibfnamefont {A.}~\bibnamefont {Rosch}},\ }\bibfield  {title} {\bibinfo {title} {Hydrodynamic long-time tails after a quantum quench},\ }\href@noop {} {\bibfield  {journal} {\bibinfo  {journal} {Physical Review A}\ }\textbf {\bibinfo {volume} {89}},\ \bibinfo {pages} {053608} (\bibinfo {year} {2014})}\BibitemShut {NoStop}%
\bibitem [{\citenamefont {Essler}(2023)}]{essler2023short}%
  \BibitemOpen
  \bibfield  {author} {\bibinfo {author} {\bibfnamefont {F.~H.}\ \bibnamefont {Essler}},\ }\bibfield  {title} {\bibinfo {title} {A short introduction to generalized hydrodynamics},\ }\href@noop {} {\bibfield  {journal} {\bibinfo  {journal} {Physica A: Statistical Mechanics and its Applications}\ }\textbf {\bibinfo {volume} {631}},\ \bibinfo {pages} {127572} (\bibinfo {year} {2023})}\BibitemShut {NoStop}%
\bibitem [{\citenamefont {Bertini}\ \emph {et~al.}(2016)\citenamefont {Bertini}, \citenamefont {Collura}, \citenamefont {De~Nardis},\ and\ \citenamefont {Fagotti}}]{bertini2016transport}%
  \BibitemOpen
  \bibfield  {author} {\bibinfo {author} {\bibfnamefont {B.}~\bibnamefont {Bertini}}, \bibinfo {author} {\bibfnamefont {M.}~\bibnamefont {Collura}}, \bibinfo {author} {\bibfnamefont {J.}~\bibnamefont {De~Nardis}},\ and\ \bibinfo {author} {\bibfnamefont {M.}~\bibnamefont {Fagotti}},\ }\bibfield  {title} {\bibinfo {title} {Transport in out-of-equilibrium xxz chains: exact profiles of charges and currents},\ }\href@noop {} {\bibfield  {journal} {\bibinfo  {journal} {Physical review letters}\ }\textbf {\bibinfo {volume} {117}},\ \bibinfo {pages} {207201} (\bibinfo {year} {2016})}\BibitemShut {NoStop}%
\bibitem [{\citenamefont {Castro-Alvaredo}\ \emph {et~al.}(2016)\citenamefont {Castro-Alvaredo}, \citenamefont {Doyon},\ and\ \citenamefont {Yoshimura}}]{castro2016emergent}%
  \BibitemOpen
  \bibfield  {author} {\bibinfo {author} {\bibfnamefont {O.~A.}\ \bibnamefont {Castro-Alvaredo}}, \bibinfo {author} {\bibfnamefont {B.}~\bibnamefont {Doyon}},\ and\ \bibinfo {author} {\bibfnamefont {T.}~\bibnamefont {Yoshimura}},\ }\bibfield  {title} {\bibinfo {title} {Emergent hydrodynamics in integrable quantum systems out of equilibrium},\ }\href@noop {} {\bibfield  {journal} {\bibinfo  {journal} {Physical Review X}\ }\textbf {\bibinfo {volume} {6}},\ \bibinfo {pages} {041065} (\bibinfo {year} {2016})}\BibitemShut {NoStop}%
\bibitem [{\citenamefont {Wellnitz}\ \emph {et~al.}(2022)\citenamefont {Wellnitz}, \citenamefont {Preisser}, \citenamefont {Alba}, \citenamefont {Dubail},\ and\ \citenamefont {Schachenmayer}}]{wellnitz2022rise}%
  \BibitemOpen
  \bibfield  {author} {\bibinfo {author} {\bibfnamefont {D.}~\bibnamefont {Wellnitz}}, \bibinfo {author} {\bibfnamefont {G.}~\bibnamefont {Preisser}}, \bibinfo {author} {\bibfnamefont {V.}~\bibnamefont {Alba}}, \bibinfo {author} {\bibfnamefont {J.}~\bibnamefont {Dubail}},\ and\ \bibinfo {author} {\bibfnamefont {J.}~\bibnamefont {Schachenmayer}},\ }\bibfield  {title} {\bibinfo {title} {Rise and fall, and slow rise again, of operator entanglement under dephasing},\ }\href@noop {} {\bibfield  {journal} {\bibinfo  {journal} {Physical Review Letters}\ }\textbf {\bibinfo {volume} {129}},\ \bibinfo {pages} {170401} (\bibinfo {year} {2022})}\BibitemShut {NoStop}%
\bibitem [{\citenamefont {Preisser}\ \emph {et~al.}(2023)\citenamefont {Preisser}, \citenamefont {Wellnitz}, \citenamefont {Botzung},\ and\ \citenamefont {Schachenmayer}}]{preisser2023comparing}%
  \BibitemOpen
  \bibfield  {author} {\bibinfo {author} {\bibfnamefont {G.}~\bibnamefont {Preisser}}, \bibinfo {author} {\bibfnamefont {D.}~\bibnamefont {Wellnitz}}, \bibinfo {author} {\bibfnamefont {T.}~\bibnamefont {Botzung}},\ and\ \bibinfo {author} {\bibfnamefont {J.}~\bibnamefont {Schachenmayer}},\ }\bibfield  {title} {\bibinfo {title} {Comparing bipartite entropy growth in open-system matrix-product simulation methods},\ }\href@noop {} {\bibfield  {journal} {\bibinfo  {journal} {Physical Review A}\ }\textbf {\bibinfo {volume} {108}},\ \bibinfo {pages} {012616} (\bibinfo {year} {2023})}\BibitemShut {NoStop}%
\bibitem [{\citenamefont {Ronzheimer}\ \emph {et~al.}(2013)\citenamefont {Ronzheimer}, \citenamefont {Schreiber}, \citenamefont {Braun}, \citenamefont {Hodgman}, \citenamefont {Langer}, \citenamefont {McCulloch}, \citenamefont {Heidrich-Meisner}, \citenamefont {Bloch},\ and\ \citenamefont {Schneider}}]{ronzheimer2013expansion}%
  \BibitemOpen
  \bibfield  {author} {\bibinfo {author} {\bibfnamefont {J.~P.}\ \bibnamefont {Ronzheimer}}, \bibinfo {author} {\bibfnamefont {M.}~\bibnamefont {Schreiber}}, \bibinfo {author} {\bibfnamefont {S.}~\bibnamefont {Braun}}, \bibinfo {author} {\bibfnamefont {S.~S.}\ \bibnamefont {Hodgman}}, \bibinfo {author} {\bibfnamefont {S.}~\bibnamefont {Langer}}, \bibinfo {author} {\bibfnamefont {I.~P.}\ \bibnamefont {McCulloch}}, \bibinfo {author} {\bibfnamefont {F.}~\bibnamefont {Heidrich-Meisner}}, \bibinfo {author} {\bibfnamefont {I.}~\bibnamefont {Bloch}},\ and\ \bibinfo {author} {\bibfnamefont {U.}~\bibnamefont {Schneider}},\ }\bibfield  {title} {\bibinfo {title} {Expansion dynamics of interacting bosons in homogeneous lattices in one and two dimensions},\ }\href@noop {} {\bibfield  {journal} {\bibinfo  {journal} {Physical review letters}\ }\textbf {\bibinfo {volume} {110}},\ \bibinfo {pages} {205301} (\bibinfo {year} {2013})}\BibitemShut {NoStop}%
\bibitem [{\citenamefont {Vidmar}\ \emph {et~al.}(2015)\citenamefont {Vidmar}, \citenamefont {Ronzheimer}, \citenamefont {Schreiber}, \citenamefont {Braun}, \citenamefont {Hodgman}, \citenamefont {Langer}, \citenamefont {Heidrich-Meisner}, \citenamefont {Bloch},\ and\ \citenamefont {Schneider}}]{vidmar2015dynamical}%
  \BibitemOpen
  \bibfield  {author} {\bibinfo {author} {\bibfnamefont {L.}~\bibnamefont {Vidmar}}, \bibinfo {author} {\bibfnamefont {J.~P.}\ \bibnamefont {Ronzheimer}}, \bibinfo {author} {\bibfnamefont {M.}~\bibnamefont {Schreiber}}, \bibinfo {author} {\bibfnamefont {S.}~\bibnamefont {Braun}}, \bibinfo {author} {\bibfnamefont {S.~S.}\ \bibnamefont {Hodgman}}, \bibinfo {author} {\bibfnamefont {S.}~\bibnamefont {Langer}}, \bibinfo {author} {\bibfnamefont {F.}~\bibnamefont {Heidrich-Meisner}}, \bibinfo {author} {\bibfnamefont {I.}~\bibnamefont {Bloch}},\ and\ \bibinfo {author} {\bibfnamefont {U.}~\bibnamefont {Schneider}},\ }\bibfield  {title} {\bibinfo {title} {Dynamical quasicondensation of hard-core bosons at finite momenta},\ }\href@noop {} {\bibfield  {journal} {\bibinfo  {journal} {Physical review letters}\ }\textbf {\bibinfo {volume} {115}},\ \bibinfo {pages} {175301} (\bibinfo {year} {2015})}\BibitemShut {NoStop}%
\bibitem [{\citenamefont {Alba}\ and\ \citenamefont {Carollo}(2021)}]{alba2021spreading}%
  \BibitemOpen
  \bibfield  {author} {\bibinfo {author} {\bibfnamefont {V.}~\bibnamefont {Alba}}\ and\ \bibinfo {author} {\bibfnamefont {F.}~\bibnamefont {Carollo}},\ }\bibfield  {title} {\bibinfo {title} {Spreading of correlations in markovian open quantum systems},\ }\href@noop {} {\bibfield  {journal} {\bibinfo  {journal} {Physical Review B}\ }\textbf {\bibinfo {volume} {103}},\ \bibinfo {pages} {L020302} (\bibinfo {year} {2021})}\BibitemShut {NoStop}%
\bibitem [{\citenamefont {Carollo}\ and\ \citenamefont {Alba}(2022)}]{carollo2022dissipative}%
  \BibitemOpen
  \bibfield  {author} {\bibinfo {author} {\bibfnamefont {F.}~\bibnamefont {Carollo}}\ and\ \bibinfo {author} {\bibfnamefont {V.}~\bibnamefont {Alba}},\ }\bibfield  {title} {\bibinfo {title} {Dissipative quasiparticle picture for quadratic markovian open quantum systems},\ }\href@noop {} {\bibfield  {journal} {\bibinfo  {journal} {Physical Review B}\ }\textbf {\bibinfo {volume} {105}},\ \bibinfo {pages} {144305} (\bibinfo {year} {2022})}\BibitemShut {NoStop}%
\bibitem [{\citenamefont {Alba}\ and\ \citenamefont {Carollo}(2023)}]{alba2023logarithmic}%
  \BibitemOpen
  \bibfield  {author} {\bibinfo {author} {\bibfnamefont {V.}~\bibnamefont {Alba}}\ and\ \bibinfo {author} {\bibfnamefont {F.}~\bibnamefont {Carollo}},\ }\bibfield  {title} {\bibinfo {title} {Logarithmic negativity in out-of-equilibrium open free-fermion chains: An exactly solvable case},\ }\href@noop {} {\bibfield  {journal} {\bibinfo  {journal} {SciPost Physics}\ }\textbf {\bibinfo {volume} {15}},\ \bibinfo {pages} {124} (\bibinfo {year} {2023})}\BibitemShut {NoStop}%
\bibitem [{\citenamefont {Vidal}(2004)}]{vidal2004TEBD}%
  \BibitemOpen
  \bibfield  {author} {\bibinfo {author} {\bibfnamefont {G.}~\bibnamefont {Vidal}},\ }\bibfield  {title} {\bibinfo {title} {Efficient simulation of one-dimensional quantum many-body systems},\ }\href@noop {} {\bibfield  {journal} {\bibinfo  {journal} {Phys. Rev. Lett.}\ }\textbf {\bibinfo {volume} {93}},\ \bibinfo {pages} {040502} (\bibinfo {year} {2004})}\BibitemShut {NoStop}%
\bibitem [{\citenamefont {White}\ and\ \citenamefont {Feiguin}(2004)}]{White2004_tDMRG}%
  \BibitemOpen
  \bibfield  {author} {\bibinfo {author} {\bibfnamefont {S.~R.}\ \bibnamefont {White}}\ and\ \bibinfo {author} {\bibfnamefont {A.~E.}\ \bibnamefont {Feiguin}},\ }\bibfield  {title} {\bibinfo {title} {Real-time evolution using the density matrix renormalization group},\ }\href@noop {} {\bibfield  {journal} {\bibinfo  {journal} {Physical Review Letters}\ }\textbf {\bibinfo {volume} {93}} (\bibinfo {year} {2004})}\BibitemShut {NoStop}%
\bibitem [{\citenamefont {Werner}\ \emph {et~al.}(2016)\citenamefont {Werner}, \citenamefont {Jaschke}, \citenamefont {Silvi}, \citenamefont {Kliesch}, \citenamefont {Calarco}, \citenamefont {Eisert},\ and\ \citenamefont {Montangero}}]{werner2016lpdo}%
  \BibitemOpen
  \bibfield  {author} {\bibinfo {author} {\bibfnamefont {A.~H.}\ \bibnamefont {Werner}}, \bibinfo {author} {\bibfnamefont {D.}~\bibnamefont {Jaschke}}, \bibinfo {author} {\bibfnamefont {P.}~\bibnamefont {Silvi}}, \bibinfo {author} {\bibfnamefont {M.}~\bibnamefont {Kliesch}}, \bibinfo {author} {\bibfnamefont {T.}~\bibnamefont {Calarco}}, \bibinfo {author} {\bibfnamefont {J.}~\bibnamefont {Eisert}},\ and\ \bibinfo {author} {\bibfnamefont {S.}~\bibnamefont {Montangero}},\ }\bibfield  {title} {\bibinfo {title} {Positive tensor network approach for simulating open quantum many-body systems},\ }\href@noop {} {\bibfield  {journal} {\bibinfo  {journal} {Phys. Rev. Lett.}\ }\textbf {\bibinfo {volume} {116}},\ \bibinfo {pages} {237201} (\bibinfo {year} {2016})}\BibitemShut {NoStop}%
\bibitem [{\citenamefont {Lumia}\ \emph {et~al.}(2024)\citenamefont {Lumia}, \citenamefont {Tirrito}, \citenamefont {Fazio},\ and\ \citenamefont {Collura}}]{lumia2024}%
  \BibitemOpen
  \bibfield  {author} {\bibinfo {author} {\bibfnamefont {L.}~\bibnamefont {Lumia}}, \bibinfo {author} {\bibfnamefont {E.}~\bibnamefont {Tirrito}}, \bibinfo {author} {\bibfnamefont {R.}~\bibnamefont {Fazio}},\ and\ \bibinfo {author} {\bibfnamefont {M.}~\bibnamefont {Collura}},\ }\bibfield  {title} {\bibinfo {title} {Measurement-induced transitions beyond gaussianity: A single particle description},\ }\href@noop {} {\bibfield  {journal} {\bibinfo  {journal} {Phys. Rev. Res.}\ }\textbf {\bibinfo {volume} {6}},\ \bibinfo {pages} {023176} (\bibinfo {year} {2024})}\BibitemShut {NoStop}%
\bibitem [{\citenamefont {Genoni}\ \emph {et~al.}(2008)\citenamefont {Genoni}, \citenamefont {Paris},\ and\ \citenamefont {Banaszek}}]{Genoni2008}%
  \BibitemOpen
  \bibfield  {author} {\bibinfo {author} {\bibfnamefont {M.~G.}\ \bibnamefont {Genoni}}, \bibinfo {author} {\bibfnamefont {M.~G.~A.}\ \bibnamefont {Paris}},\ and\ \bibinfo {author} {\bibfnamefont {K.}~\bibnamefont {Banaszek}},\ }\bibfield  {title} {\bibinfo {title} {Quantifying the non-gaussian character of a quantum state by quantum relative entropy},\ }\href@noop {} {\bibfield  {journal} {\bibinfo  {journal} {Phys. Rev. A}\ }\textbf {\bibinfo {volume} {78}},\ \bibinfo {pages} {060303} (\bibinfo {year} {2008})}\BibitemShut {NoStop}%
\bibitem [{\citenamefont {Genoni}\ and\ \citenamefont {Paris}(2010)}]{Genoni2010}%
  \BibitemOpen
  \bibfield  {author} {\bibinfo {author} {\bibfnamefont {M.~G.}\ \bibnamefont {Genoni}}\ and\ \bibinfo {author} {\bibfnamefont {M.~G.~A.}\ \bibnamefont {Paris}},\ }\bibfield  {title} {\bibinfo {title} {Quantifying non-gaussianity for quantum information},\ }\href@noop {} {\bibfield  {journal} {\bibinfo  {journal} {Phys. Rev. A}\ }\textbf {\bibinfo {volume} {82}},\ \bibinfo {pages} {052341} (\bibinfo {year} {2010})}\BibitemShut {NoStop}%
\bibitem [{\citenamefont {Vedral}(2002)}]{vedral2002relativeentropy}%
  \BibitemOpen
  \bibfield  {author} {\bibinfo {author} {\bibfnamefont {V.}~\bibnamefont {Vedral}},\ }\bibfield  {title} {\bibinfo {title} {The role of relative entropy in quantum information theory},\ }\href@noop {} {\bibfield  {journal} {\bibinfo  {journal} {Rev. Mod. Phys.}\ }\textbf {\bibinfo {volume} {74}},\ \bibinfo {pages} {197} (\bibinfo {year} {2002})}\BibitemShut {NoStop}%
\bibitem [{\citenamefont {Marian}\ and\ \citenamefont {Marian}(2013)}]{Marian2013}%
  \BibitemOpen
  \bibfield  {author} {\bibinfo {author} {\bibfnamefont {P.}~\bibnamefont {Marian}}\ and\ \bibinfo {author} {\bibfnamefont {T.~A.}\ \bibnamefont {Marian}},\ }\bibfield  {title} {\bibinfo {title} {Relative entropy is an exact measure of non-gaussianity},\ }\href@noop {} {\bibfield  {journal} {\bibinfo  {journal} {Phys. Rev. A}\ }\textbf {\bibinfo {volume} {88}},\ \bibinfo {pages} {012322} (\bibinfo {year} {2013})}\BibitemShut {NoStop}%
\bibitem [{Note1()}]{Note1}%
  \BibitemOpen
  \bibinfo {note} {Notice that the replacement ${\protect \rm tr}(\protect \hat \rho \log \protect \hat \rho _{GGE}) \to {\protect \rm tr}(\protect \hat \rho _{GGE}\log \protect \hat \rho _{GGE})$ is not exact for $S(\protect \hat \rho ||\protect \hat \rho _{GGE})$, because $\protect \hat \rho $ and $\protect \hat \rho _{GGE}$ have different two-point functions. Nevertheless, the error that is made in the replacement is under control and we can still make it within our approximations: as we show in the Appendix \ref {A: non-gaussianity}, the error is of same the order of the distance between $\protect \hat \rho _G$ and $\protect \hat \rho _{GGE}$, which is a small correction to the distance between $\protect \hat \rho $ and $\protect \hat \rho _{GGE}$ if the largest component of the error is brought by non-Gaussianity, as we expect.}\BibitemShut {Stop}%
\bibitem [{\citenamefont {Peschel}\ and\ \citenamefont {Eisler}(2009)}]{peschel2009reduced}%
  \BibitemOpen
  \bibfield  {author} {\bibinfo {author} {\bibfnamefont {I.}~\bibnamefont {Peschel}}\ and\ \bibinfo {author} {\bibfnamefont {V.}~\bibnamefont {Eisler}},\ }\bibfield  {title} {\bibinfo {title} {Reduced density matrices and entanglement entropy in free lattice models},\ }\href@noop {} {\bibfield  {journal} {\bibinfo  {journal} {Journal of physics a: mathematical and theoretical}\ }\textbf {\bibinfo {volume} {42}},\ \bibinfo {pages} {504003} (\bibinfo {year} {2009})}\BibitemShut {NoStop}%
\bibitem [{\citenamefont {Surace}\ and\ \citenamefont {Tagliacozzo}(2022)}]{surace2022}%
  \BibitemOpen
  \bibfield  {author} {\bibinfo {author} {\bibfnamefont {J.}~\bibnamefont {Surace}}\ and\ \bibinfo {author} {\bibfnamefont {L.}~\bibnamefont {Tagliacozzo}},\ }\bibfield  {title} {\bibinfo {title} {{Fermionic Gaussian states: an introduction to numerical approaches}},\ }\href@noop {} {\bibfield  {journal} {\bibinfo  {journal} {SciPost Phys. Lect. Notes}\ ,\ \bibinfo {pages} {54}} (\bibinfo {year} {2022})}\BibitemShut {NoStop}%
\bibitem [{\citenamefont {Antal}\ \emph {et~al.}(1999)\citenamefont {Antal}, \citenamefont {R\'acz}, \citenamefont {R\'akos},\ and\ \citenamefont {Sch\"utz}}]{antal1999profile}%
  \BibitemOpen
  \bibfield  {author} {\bibinfo {author} {\bibfnamefont {T.}~\bibnamefont {Antal}}, \bibinfo {author} {\bibfnamefont {Z.}~\bibnamefont {R\'acz}}, \bibinfo {author} {\bibfnamefont {A.}~\bibnamefont {R\'akos}},\ and\ \bibinfo {author} {\bibfnamefont {G.~M.}\ \bibnamefont {Sch\"utz}},\ }\bibfield  {title} {\bibinfo {title} {Transport in the $\mathrm{XX}$ chain at zero temperature: Emergence of flat magnetization profiles},\ }\href@noop {} {\bibfield  {journal} {\bibinfo  {journal} {Phys. Rev. E}\ }\textbf {\bibinfo {volume} {59}},\ \bibinfo {pages} {4912} (\bibinfo {year} {1999})}\BibitemShut {NoStop}%
\bibitem [{\citenamefont {Grafakos}(2010)}]{GrafakosClassicalFourierAnalysis}%
  \BibitemOpen
  \bibfield  {author} {\bibinfo {author} {\bibfnamefont {L.}~\bibnamefont {Grafakos}},\ }\href@noop {} {\emph {\bibinfo {title} {Classical Fourier Analysis}}}\ (\bibinfo  {publisher} {Springer NY},\ \bibinfo {year} {2010})\BibitemShut {NoStop}%
\end{thebibliography}%

\end{document}